\documentclass[%
	aps,%
	prb,%
	twocolumn,%
	10pt,%
	balancelastpage,%
	amsmath,%
	amssymb,%
	superscriptaddress,%
	floatfix,%
	footinbib]{revtex4-2}

\usepackage[utf8]{inputenc}
\usepackage[T1]{fontenc}
\usepackage{lmodern}
\usepackage{graphicx}
\usepackage{overpic}
\usepackage{float}
\usepackage{subfig}
\usepackage{xcolor}
\usepackage{subcaption}
\usepackage{amssymb,amsmath,amsfonts,bm}
\usepackage{braket}
\usepackage[colorlinks=true,citecolor=blue,linkcolor=blue,urlcolor=blue]{hyperref}
\usepackage{adjustbox}

\usepackage{booktabs}
\usepackage[table]{xcolor}
\usepackage{adjustbox}

\definecolor{HeaderBlue}{RGB}{225,235,248}
\definecolor{FirstColBlue}{RGB}{238,244,252}
\definecolor{EmptyGray}{RGB}{245,245,245}

\newcommand{\emptycell}{\cellcolor{EmptyGray}}

\newcommand{\ud}{\mathrm{d}}

\newcommand{\Tr}{\operatorname{Tr}}

\begin{document}

\title{{Semiclassical Langevin dynamics of long-range dissipative time crystals}}

\author{Reyhaneh Khasseh}
\affiliation{Theoretical Physics III, Center for Electronic Correlations and Magnetism,\\
Institute of Physics, University of Augsburg, D-86135 Augsburg, Germany}
\author{Rosario Fazio}
\affiliation{Dipartimento di Fisica, Universit\`a di Napoli ``Federico II'', Monte S. Angelo, I-80126 Napoli, Italy}
\affiliation{The Abdus Salam International Centre for Theoretical Physics, Strada Costiera 11, I-34151 Trieste, Italy}
\author{Angelo Russomanno}
\affiliation{Dipartimento di Fisica, Universit\`a di Napoli ``Federico II'', Monte S. Angelo, I-80126 Napoli, Italy}
\date{\today}
\begin{abstract}
We develop a {semiclassical Langevin} approach to study finite-size effects in dissipative time-crystalline spin systems. For a spin-$1/2$ model with power-law Lindblad operators, we find finite-size oscillations exponentially decaying in time, and their decay rate decreases algebraically with system size in the long-range regime. The scaling exponent of this decay time provides a direct diagnostic of the robustness of the time-crystalline, and we find that this robustness extends beyond the range of power-law dissipation exponents where the system dynamics is mean-field in the thermodynamic limit. We also apply our approach to a spin-one model with local dissipation and long-range Hamiltonian interactions, {formulating it} in terms of Gell-Mann variables. In this case, finite-size stability is quantified through the deviation time from the mean-field behavior and the behavior of the dominant Fourier peak. We find that both quantities scale as a power law with the system size -- marking thereby a time-crystal behavior -- when the exponent of the power-law interaction is below a certain threshold, that turns out to be in agreement with previous cumulant-expansion findings. Our results show that semiclassical Langevin dynamics provides a useful finite-size probe of dissipative time crystals in regimes beyond exact Lindblad simulations.
%
\end{abstract}

\maketitle

%
\section{Introduction}

Time crystals extend the notion of spontaneous symmetry breaking to the time domain. The original proposal of a quantum time crystal suggested that a many-body system could display persistent motion in its lowest-energy state~\cite{Wilczek2012,Shapere2012}. Subsequent no-go theorems showed that such behavior is forbidden in equilibrium ground states and thermal states of generic closed quantum systems~\cite{Bruno2013NoGo,Watanabe2015}. These results shifted the search for time-crystalline order toward nonequilibrium settings, where external driving, dissipation, long-range interactions, or coupling to an environment can stabilize persistent oscillatory dynamics.


Open quantum systems provide a particularly natural route to continuous time-crystalline behavior. In driven-dissipative systems, the competition between coherent evolution and dissipation can prevent relaxation to a time-independent state and instead generate stable collective oscillations {that become persistent for an infinite time in the thermodynamic limit, thereby breaking the continuous (or sometimes discrete) time translation symmetry. This phenomenon occurs in the so called quantum dissipative (or boundary) time crystals (BTC) that have been object of intense theoretical}~\cite{dj21-gmdj,Delmonte_2023,PhysRevA.101.033839,PhysRevA.110.L010202,PhysRevResearch.6.033185,PRXQuantum.5.030325,PhysRevLett.120.040404,Zhu_TC_2019,PhysRevA.108.L041303,osullivan2019dissipativediscretetimecrystals,harmon2025collectiveexcitationsdissipativetime,Alaeian_2022,Buca2019,Booker2020,Carollo2022,iemini-prl-2018:boundary-time-crystals,gong_2018,
Tucker2018, Shammah2018,Zhu_2019, lledo-pra-2019,
carollo-prl-2020:quantum-engine-time-translation-symmetry-breaking,riera-campeny-2019,seibold-pra-2020,iemini-prb-2021:btc-collective-d-level-systems,
bonaiuto-prl-2021:correlations-emitter-waveguide,piccitto-prb-2021:generalized-spin-models,hajdusek_2022,sarkar_2022,krishna_2022,dc2s-94gv,Wang2025BoundaryTimeCrystals,PhysRevA.99.053605,PhysRevLett.115.163601} {and experimental research~\cite{chen2023inherent,Liu2025Bifurcation,K_Science22,PhysRevLett.127.043602,Wu_2024,Lai_2025} in recent years. In all these cases time-translation symmetry is spontaneously broken in the thermodynamic limit:} In finite systems, the collective oscillations may decay at long times, but the defining feature of the BTC phase is that the lifetime diverges with system size, providing persistent oscillations in the thermodynamic limit.


The paradigmatic BTC model involves spin-$1/2$ particles subject to coherent driving and collective dissipation~\cite{iemini-prl-2018:boundary-time-crystals}. In the thermodynamic limit, for perfectly collective models where interactions are infinite-range, the dynamics can be described by mean-field equations for the global magnetization, which display stable closed-orbit~\cite{iemini-prl-2018:boundary-time-crystals} or limit-cycle~\cite{PhysRevLett.132.183803,b6gq-z5nf} behavior. However, realistic systems are rarely perfectly collective. A too strong local dissipation and finite-range effects may destroy the collective oscillation and restore relaxation to a stationary state~\cite{Tucker2018,Shammah2018,Wang2025BoundaryTimeCrystals}. This motivates the study of how robust BTC behavior remains when the ideal collective structure is weakened, and there is indeed a range of parameters where this robustness is attained, a crucial finding for experimental observation~\cite{K_Science22}.

{While most of the work on time crystals
in open systems has been performed on infinite range systems and that only very few papers addressed finite range cases, there are some cases that consider a different perspective. 
One example is the spin-1/2 model of }Ref.~\cite{Passarelli2022LongRangeLindbladians}, where the collective Lindblad operators of~\cite{iemini-prl-2018:boundary-time-crystals} were replaced by power-law long-range Lindblad operators. In that model, the range of the dissipative coupling is controlled by an exponent ${\alpha}$. It was shown that persistent BTC oscillations survive when the dissipative coupling is sufficiently long ranged, in particular for ${\alpha}<1$ and dissipation is weak enough, due to the equivalence to the mean field in the thermodynamic limit for ${\alpha} < 1$. This demonstrates that exact collective spin symmetry is not strictly necessary for dissipative time-crystalline order, provided that the Lindbladian remains sufficiently long ranged. 

{Another example is provided by~\cite{Wang2025BoundaryTimeCrystals}, where a spin-1 system with power-law long-range interacting Hamiltonian behaves as a dissipative time crystal in presence of a local dissipation. It is important to emphasize that the dissipation is strictly local, in contrast to the collective dissipation of most proposals}. This model displays time-crystal oscillations in the fully-connected case that persist when the interaction is long-range, and disappears abruptly when the power-law-decay exponent goes beyond a threshold between 1 and 1.5, as a cumulant-expansion analysis shows. {Quite remarkably, the time-crystal behavior persists beyond the threshold $\alpha=1$, above which the thermodynamic-limit dynamics can be proved to be not described by mean-field equations.}

In this work, we revisit both these models using a {semiclassical Langevin} approach. This is a semiclassical method rooted in the truncated-Wigner approximation methods. These methods are based on the representation of the wavefunction as a Wigner function, that can be constructed on a continuous phase space~\cite{Polkovnikov2010}, or on a discrete one, as appropriate for spin systems~\cite{WOOTTERS19871}. These representations amount to an expansion in a many-body operator basis, and these operators evolve according to the Heisenberg dynamics. The truncated-Wigner approximation (TWA) schemes consist in neglecting the quantum correlations between different sites and letting evolve these operators in a mean-field approximation. Quantum fluctuations are partially included as a Monte-Carlo trajectory sampling over different classical initial conditions, to represent the weights on the different basis operators of the initial state. When the sampling is on a discrete phase space, the truncated Wigner approximation is a discrete one (DTWA) that was first introduced in the case of spin 1/2 systems~\cite{Schachenmayer2015}, and was later extended to spins of generic size (generalized DTWA), using as expansion basis the tensor products of the Gell-Mann matrices~\cite{Zhu_2019}. DTWA methods work very well in systems with long-range interactions~\cite{Czischek_2018,Kunimi2021}, and have been successfully applied in many contexts~\cite{Lepoutre_2019,Fersterer_2019,PhysRevResearch.2.023050,Schachenmayer_2015,Sundar_2019,Orioli_2018,PhysRevB.102.014303,Signoles_2021,Pucci_2016,Acevedo_2017,Covey_2018,Khasseh2021Fragility}.



These TWA methods were extended to the case of dissipative systems. One example is the Langevin treatment of the Fokker-Planck equation governing the evolution of the Wigner function on a continuous phase space~\cite{PhysRevResearch.4.043136,SciPostPhys.10.2.045,SciPostPhys.15.6.233,Tebbenjohanns_2024}, that was recently applied to a time-crystal context~\cite{b6gq-z5nf}. Also DTWA methods have been extended to the case of dissipative dynamics~\cite{Qu_2019,Liu_2020,Singh_2022}. One possible approach starts from the quantum Langevin equation~\cite{GardinerZoller}, and amounts to describing the stochastic operators as independent Wiener stochastic processes, and to applying the DTWA -- that's to say the mean-field approximation to the Heisenberg evolution of the operators of the discrete expansion basis. In this way the dynamics is approximated as an average over different realizations of the evolution of nonlinear coupled stochastic differential equations~\cite{1wwv-k7hg,PhysRevA-105-013716}. As in the usual DTWA, there is also a sampling on classical initial conditions that takes into account part of the quantum fluctuations.

We apply this Langevin DTWA method to the long-range continuous time crystals described in~\cite{Passarelli2022LongRangeLindbladians,Wang2025BoundaryTimeCrystals}. While for the spin 1/2 case~\cite{Passarelli2022LongRangeLindbladians} we can directly apply the scheme introduced in~\cite{PhysRevA-105-013716}, for the spin 1 model~\cite{Wang2025BoundaryTimeCrystals} we have to represent the dynamics on the basis of the tensor products of Gell-Mann matrices, as explained in~\cite{Zhu_2019}. For both cases we find that the sampling on initial conditions is unnecessary: Memory of the initial state is lost after a transient much shorter than the period of the time-crystal oscillations we are interested in. For the models we study the Langevin DTWA is therefore equivalent to a {semiclassical Langevin dynamics where no sampling on the initial conditions is performed.}

Our main diagnostic for the spin-$1/2$ model is the finite-size decay of the oscillation envelope. For finite systems, the magnetization oscillations decay exponentially at long times. We extract the corresponding decay rate and study its scaling with system size. In the long-range regime, the decay rate decreases algebraically with system size, providing a direct finite-size signature of BTC order. As the power-law exponent ${\alpha}$ is increased, the scaling weakens and eventually becomes compatible with the loss of persistent oscillations. This happens for $\alpha\simeq 1.2$, a value slightly above ${\alpha}=1$, the threshold below which the thermodynamic-limit behavior is a mean-field boundary time crystal~\cite{Passarelli2022LongRangeLindbladians}, as a cumulant expansion shows. Beyond $\alpha=1$ the mean-field analysis predicts relaxation to a fixed point, but  in this range quantum fluctuations are relevant in the thermodynamic limit, and a mean-field analysis is not enough~\cite{Passarelli2022LongRangeLindbladians}. Our approach provides indeed a result different from the mean-field analysis: A time-crystal behavior in a range of $\alpha>1$. This emphasizes that we take into account at least part of the quantum fluctuations, as all DTWA-related methods do. 
%

In the spin-1 case, oscillations are not symmetric around zero, and we quantify their stability using two probes: The system-size scaling of the deviation time from the mean-field trajectory and the system-size scaling of the dominant Fourier peak. We find that both quantities scale as a power law with the system size: There is a time-crystal behavior when the scaling exponent we obtain is positive. Both approaches provide as a threshold for the time-crystal behavior the value of the power-law interaction exponent $\alpha \simeq 1.2$. Although for the deviation-time analysis this is probably an artifact -- being the method rigorously valid only for $\alpha \leq 1$ where the thermodynamic-limit behavior is described by the mean-field dynamics -- this result is in full agreement with the cumulant-expansion analysis of~\cite{Wang2025BoundaryTimeCrystals}, where this threshold was set at a value of $\alpha$ in the interval $[1,1.5]$.
Beyond that, our analysis provides the interesting information that both the deviation time and the dominant Fourier peak scale with an exponent that depends nontrivially on $\alpha$, also in the range $\alpha\in[0,1]$ where the thermodynamic-limit behavior is mean-field. 
Therefore a mean-field behavior is valid for the thermodynamic limit but not for the system-size scaling exponent of the finite-size transient.

The paper is organized as follows. In Sec.~\ref{sec:method} we introduce the general Langevin formulation and the numerical method used throughout the work. In Sec.~\ref{sec:spinhalf_model} we derive the semiclassical stochastic equations for the spin-$1/2$ model with power-law Lindblad operators. In Sec.~\ref{numerical:sec} we present the finite-size scaling analysis of the spin-$1/2$ dynamics and extract the decay-rate exponent of the oscillations. In Sec.~\ref{spin-one:sec} we introduce the spin-one model with local dissipation and long-range Hamiltonian interactions and derive its stochastic equations in the Gell-Mann representation. In Sec.~\ref{numerical1:sec} we analyze the spin-one dynamics using the deviation time from mean-field behavior. In Sec~\ref{conc:sec} we draw our conclusions. Finally, the appendices contain the derivations of the {semiclassical Langevin} equations (Appendixes~\ref{app:spinhalf_langevin_derivation} ~\ref{app:spinone_langevin_derivation}), benchmarks against exact Lindblad dynamics (Appendix~\ref{supp:s1}), the algebraic relations of the Gell-Mann basis (Appendix~\ref{app:gellmann-algebra}), and the analysis with the Lindblad-Heisenberg equations for collective Gell-Mann matrices of the infinite-range spin-1 case, showing the exact validity of the Lindblad mean-field equations in the thermodynamic limit (Appendix~\ref{infinite:app}).

\section{Semiclassical Langevin method}
\label{sec:method}

We consider open quantum spin systems whose density matrix evolves according to a Lindblad master equation,
\begin{equation}
  \dot{\rho}
  =
  -i[\hat{\mathcal{H}},\rho]
  +\frac{\gamma}{2}
  \sum_{j}
  \left(
  2 {\hat{L}_j} \rho {\hat{L}_j}^\dagger
  - {\hat{L}_j}^\dagger {\hat{L}_j} \rho
  - \rho {\hat{L}_j}^\dagger {\hat{L}_j}
  \right) .
  \label{eq:lindblad_master}
\end{equation}
Here $\hat{\mathcal{H}}$ is the Hamiltonian, ${\hat{L}_j}$ are dimensionless Lindblad operators, and $\gamma$ sets the overall dissipative rate. Instead of solving Eq.~\eqref{eq:lindblad_master} directly, we formulate the dynamics in terms of the corresponding quantum Langevin equation~\cite{GardinerZoller}. For a system operator $\hat{\hat{O}}$, one obtains
\begin{equation}
\label{eq:QLE}
\begin{split}
\frac{d}{dt}\hat{O}
=
i[\hat{\mathcal{H}},\hat{O}]
&-
\sum_j
\Bigg\{
[\hat{O},{\hat{L}_j}^\dagger]
\left(
\frac{\gamma}{2}{\hat{L}_j}
+
\sqrt{\gamma}\,\hat F_j(t)
\right)
\\
&\hspace{1.8cm}
+
\left(
\frac{\gamma}{2}{\hat{L}_j}^\dagger
+
\sqrt{\gamma}\,\hat F_j^\dagger(t)
\right)
[{\hat{L}_j},\hat{O}]
\Bigg\}.
\end{split}
\end{equation}
The operators $\hat F_j(t)$ describe Markovian quantum noise associated with the dissipative channels. They have zero mean and are delta-correlated in time. In the following sections, Eq.~\eqref{eq:QLE} is used as the starting point for deriving stochastic equations of motion for two different dissipative spin models, in the spirit of the semiclassical Langevin approach used in Ref.~\cite{PhysRevA-105-013716}.

To obtain a tractable semiclassical description, we apply Eq.~\eqref{eq:QLE} to a complete set of local spin operators and then replace operator products by products of their expectation values. For spin-$1/2$ systems, the dynamical variables are the local spin components
\begin{equation}
  s_i^\mu(t)=\langle \hat{\sigma}_i^\alpha(t)\rangle,
  \qquad
  \mu=x,y,z .
\end{equation}
For spin-one systems, we use instead the eight local Gell-Mann variables
\begin{equation}
  \lambda_{\mu i}(t)=\langle \hat{\Lambda}_{\mu i}(t)\rangle,
  \qquad
  \mu=1,\ldots,8 .
\end{equation}
This approximation neglects connected correlations beyond the level retained by the stochastic noise terms, while preserving the deterministic mean-field dynamics and the leading finite-size fluctuations generated by the dissipative channels.

The quantum noise operators are represented by real Gaussian white noises. Equivalently, for each dissipative channel one writes
\begin{equation}
  \hat F_j(t)
  =
  \frac{1}{2}
  \left[
  \xi_j^{(1)}(t)
  +
  i\xi_j^{(2)}(t)
  \right],
\end{equation}
with
\begin{equation}
  \left\langle
  \xi_j^{(\alpha)}(t)
  \xi_l^{(\beta)}(t')
  \right\rangle
  =
  \delta_{jl}
  \delta_{\alpha\beta}
  \delta(t-t') .
\end{equation}
This gives a set of stochastic differential equations for the semiclassical variables. We interpret these equations in the Itô sense.

The stochastic equations are solved numerically by discretizing time with step $\Delta t$. The deterministic drift terms are integrated using a Runge--Kutta scheme, while the stochastic terms are sampled using independent Gaussian Wiener increments satisfying
\begin{equation}
  \Delta W_j^{(\mu)}
  =
  \sqrt{\Delta t}\,\zeta_j^{(\mu)},
  \qquad
  \left\langle
  \zeta_j^{(\mu)}
  \zeta_l^{(\nu)}
  \right\rangle
  =
  \delta_{jl}\delta_{\mu\nu},
\end{equation}
where the $\zeta_j^{(\mu)}$ are independent normal random variables with zero mean and unit variance, {$j,\,l$ are the site indices, and $\mu,\,\nu$ the operator ones}. Observables are obtained by averaging over many stochastic trajectories. The order parameter used to characterize the time-crystalline dynamics is the longitudinal magnetization
\begin{equation}\label{eq:Mz_traj_avg}
  M_z(t)
  =
  \frac{1}{N_{\mathrm{traj}}}
  \sum_{r=1}^{N_{\mathrm{traj}}}
  \frac{1}{L}
  \sum_{i=1}^{L}
  s_{i,r}^z(t).
\end{equation}
Here \(r\) labels the stochastic trajectory, and the overline denotes the average over \(N_{\mathrm{traj}}\) independent realizations. For the spin-$1/2$ model, $s_i^z(t)=\langle\hat{\sigma}_i^z(t)\rangle$, while for the spin-one model it is obtained from the Gell-Mann variables. In both cases, $M_z(t)$ is averaged over stochastic trajectories and used to analyze the oscillation frequency and finite-size decay of the time-crystalline response.

Notice that here we consider one single initial condition and not perform a Monte-Carlo sampling on different initial conditions as instead it is done in the Langevin DTWA method~\cite{1wwv-k7hg,PhysRevA-105-013716}. We do that because we find no differences between the two approaches.
\section{Spin-$1/2$ model with power-law Lindblad operators}
\label{sec:spinhalf_model}

We first apply the semiclassical Langevin method to a chain of \(L\) spin-\(1/2\)
degrees of freedom. The coherent dynamics is generated by a transverse field,
\begin{equation}
  \hat{\mathcal{H}}=J\sum_{i=1}^{L}\hat{\sigma}_i^x .
\end{equation}
The coupling between different spins is not introduced through the Hamiltonian,
but rather through the dissipative part of the dynamics. Following
Ref.~\cite{Passarelli2022LongRangeLindbladians}, we consider power-law
Lindblad operators of the form
\begin{equation}
\label{eq:power-law-lindblad}
  {\hat{L}_j}=\sum_{l=1}^{L} f_{jl}({\alpha})\,\sigma_l^+,
  \qquad
  f_{jl}({\alpha})=
  \frac{K^{(L)}({\alpha})}
  {[D(|j-l|)]^{{\alpha}}}.
\end{equation}
Here
\(\sigma_l^\pm=\frac{1}{2}\left(\sigma_l^x\pm i\sigma_l^y\right)\), and
periodic boundary conditions are implemented through
\begin{equation}
  D(r)=\min(r,L-r)+1 .
\end{equation}
The Kac factor \(K^{(L)}({\alpha})\) is fixed by the normalization condition
\begin{equation}
  \sum_{j=1}^{L} f_{jl}({\alpha})=1 .
\end{equation}
The exponent \({\alpha}\) controls the range of the dissipative coupling. For
\({\alpha}=0\), all sites enter each Lindblad operator with equal weight, and the
dissipation is fully collective. Increasing \({\alpha}\) progressively suppresses
the contribution of distant sites and makes the Lindblad operators more
short-ranged.

To derive the semiclassical dynamics, we introduce local spin variables
\begin{equation}
  s_i^\alpha(t)=\langle \hat{\sigma}_i^\alpha(t)\rangle,
  \qquad
  \alpha=x,y,z .
\end{equation}
Starting from the quantum Langevin equation and factorizing connected spin
correlations at the semiclassical level, one obtains stochastic equations for
these variables. Before writing them, it is useful to absorb the spatial profile
of the Lindblad operators into an effective dissipative coupling matrix,
\begin{equation}
G_{il}({\alpha})
=
\sum_{j=1}^{L}
f_{ji}({\alpha})f_{jl}({\alpha}),
\end{equation}
and into the corresponding noise increments
\begin{equation}
d\mathcal W_i^{(\alpha)}
=
\sum_{j=1}^{L}
f_{ji}({\alpha})\,dW_j^{(\alpha)} ,
\qquad
\alpha=1,2 .
\end{equation}

\begin{figure*}
    \includegraphics[width=0.7\columnwidth]{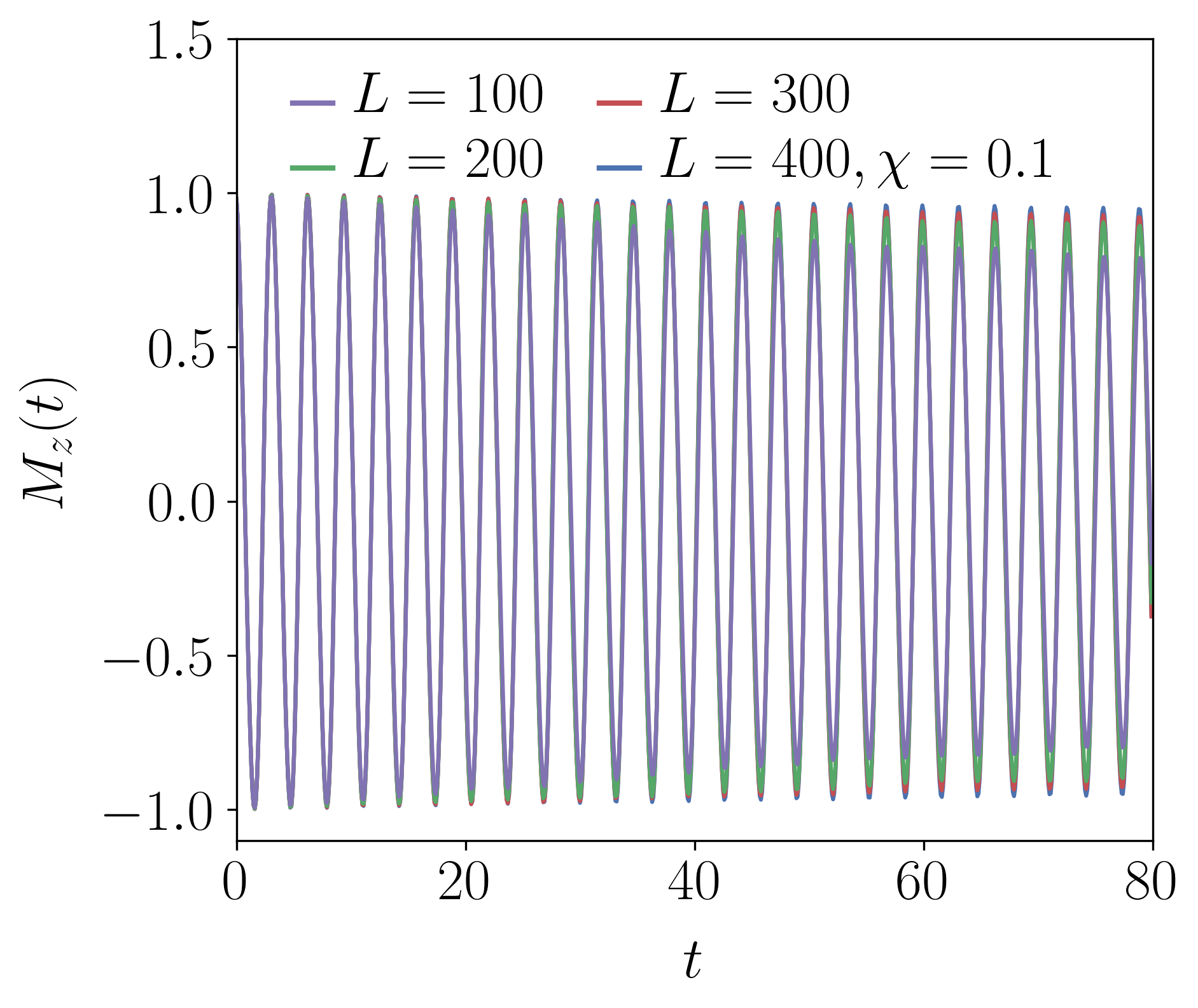}\put(-150,150){(a)}
    \includegraphics[width=0.7\columnwidth]{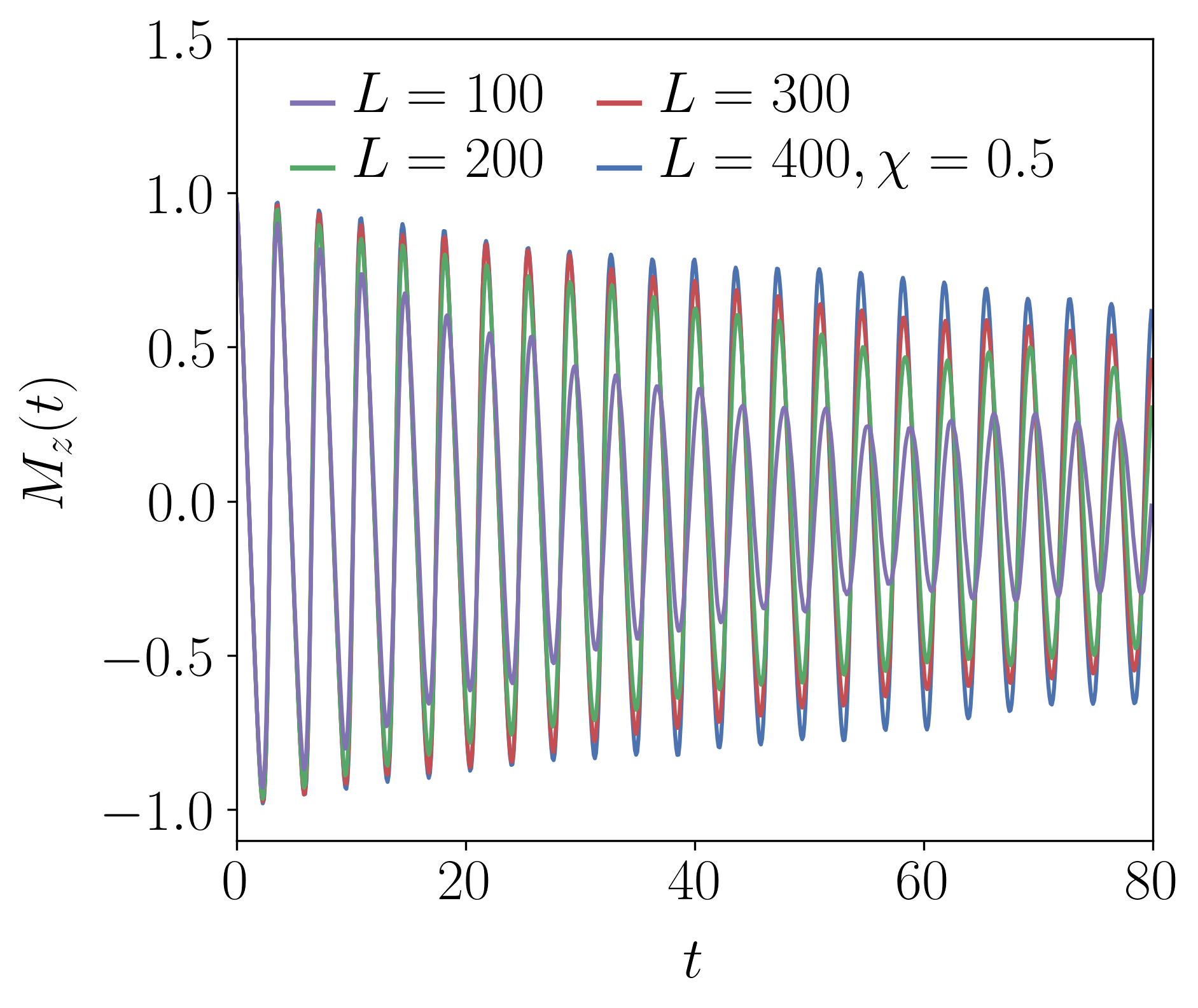}\put(-150,150){(b)}
    \includegraphics[width=0.7\columnwidth]{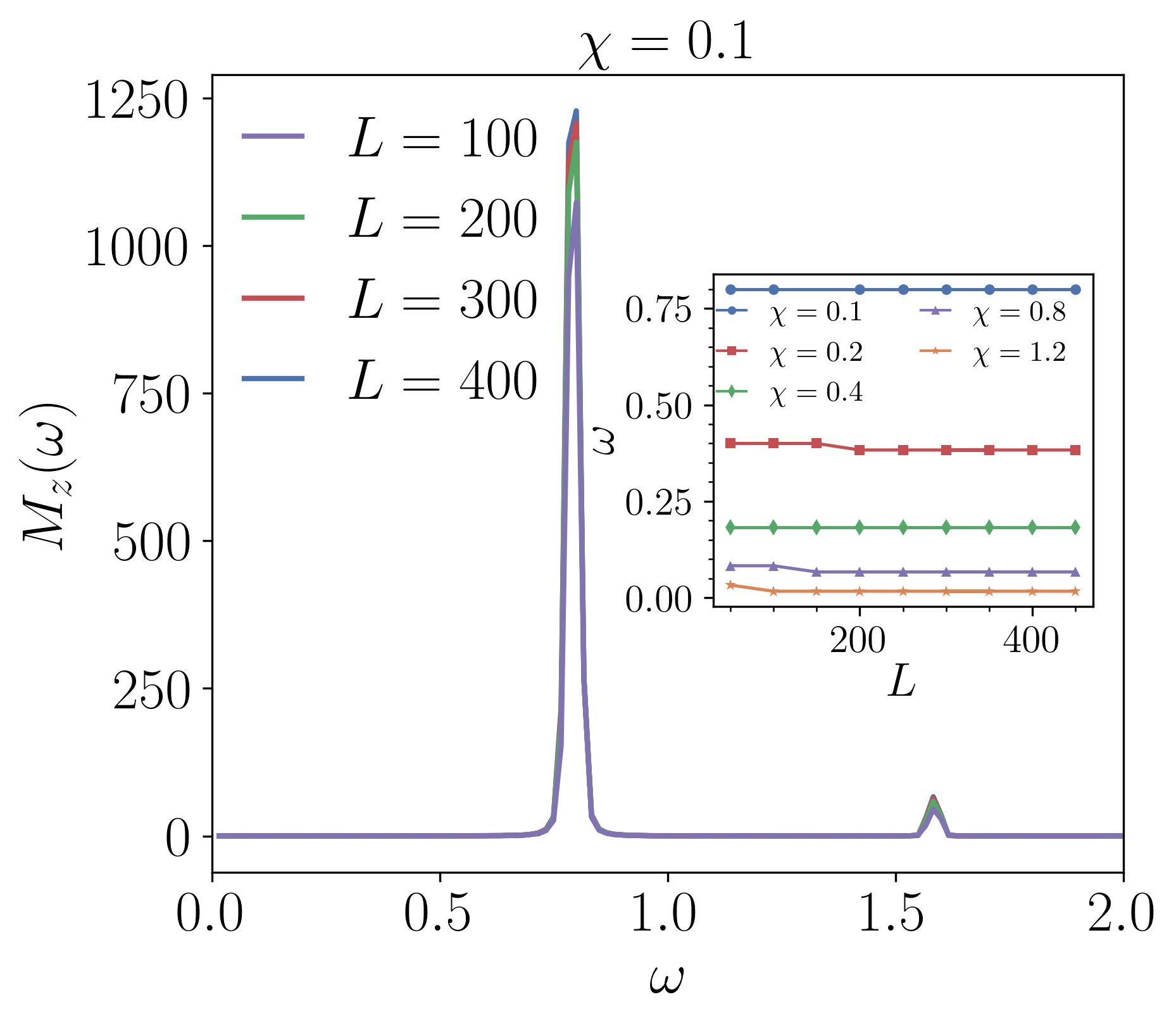}\put(-150,150){(c)}
    \caption{\textbf{Fully connected Lindblad dynamics.}
    (a),(b) Time evolution of the magnetization \(M_z(t)\) for different system sizes in the fully connected limit \({\alpha}=0\). For \(\chi<1\), the system displays oscillatory dynamics characteristic of the boundary time crystal. The damping decreases with increasing system size.
    (c) Fourier spectrum of \(M_z(t)\) for \(\chi=0.1\). A sharp dominant peak identifies a well-defined collective oscillation frequency. Inset: dominant frequency as a function of system size for different \(\chi\). The frequency decreases upon approaching the transition and vanishes in the stationary phase.}
    \label{fig:Long-range_entang}
\end{figure*}
The matrix \(G_{il}({\alpha})\) therefore measures the overlap between the
dissipative profiles associated with sites \(i\) and \(l\). In this notation the
noise increments \(d\mathcal W_i^{(\alpha)}\) are generally spatially
correlated, with correlations determined by the same matrix \(G_{il}({\alpha})\).

We also define the nonlocal transverse fields
\begin{equation}
h_i^\mu(t)
=
\sum_{l\neq i}
G_{il}({\alpha})s_l^\mu(t),
\qquad
\mu=x,y .
\end{equation}
These fields collect the contributions of all spins coupled to site \(i\)
through the dissipative channels. In terms of \(G_{il}\), \(h_i^\mu\), and
\(d\mathcal W_i^{(\alpha)}\), the semiclassical Itô equations take the compact
form
\begin{align}
d s_i^x
&=
-\frac{\gamma}{2}
\left[
G_{ii}({\alpha})s_i^x
+
s_i^z h_i^x
\right]dt
-\sqrt{\gamma}\,s_i^z\,d\mathcal W_i^{(1)},
\nonumber\\
d s_i^y
&=
\left\{
-2J s_i^z
-\frac{\gamma}{2}
\left[
G_{ii}({\alpha})s_i^y
+
s_i^z h_i^y
\right]
\right\}dt
-\sqrt{\gamma}\,s_i^z\,d\mathcal W_i^{(2)},
\nonumber\\
d s_i^z
&=
\left\{
2J s_i^y
+
\gamma G_{ii}({\alpha})(1-s_i^z)
+
\frac{\gamma}{2}
\left[
s_i^x h_i^x
+
s_i^y h_i^y
\right]
\right\}dt
\nonumber\\
&\quad
+
\sqrt{\gamma}
\left[
s_i^x\,d\mathcal W_i^{(1)}
+
s_i^y\,d\mathcal W_i^{(2)}
\right].
\label{eq:spinhalf_semiclassical_sde_compact}
\end{align}
The first terms in these equations describe the coherent precession generated by
the transverse field, while the remaining deterministic and stochastic terms are
generated by the long-range dissipative channels. The derivation of
Eq.~\eqref{eq:spinhalf_semiclassical_sde_compact} from the operator Langevin
equations is given in Appendix~\ref{app:spinhalf_langevin_derivation}.

The system is initialized with all spins polarized along the \(x\) direction, as
in the boundary-time-crystal protocol~\cite{iemini-prl-2018:boundary-time-crystals}. We use the
dimensionless dissipation strength
\begin{equation}
  \chi=\frac{\gamma}{4J}.
\end{equation}
For the fully collective case \({\alpha}=0\), the transition occurs at \(\chi=1\):
for \(\chi<1\), the thermodynamic dynamics supports persistent oscillations,
whereas for \(\chi>1\), it relaxes to a stationary magnetized state.

\section{Numerical results for the spin-$1/2$ model}\label{numerical:sec}

In this section we analyze the real-time dynamics obtained from the
semiclassical stochastic equations derived above. We focus on the
trajectory-averaged longitudinal magnetization \(M_z(t)\), defined in
Eq.~\eqref{eq:Mz_traj_avg}, and on the finite-size decay of its oscillations.
This quantity is used to extract both the oscillation frequency and the
finite-size decay rate.
and on the finite-size decay of its oscillations.

\subsection{Magnetization dynamics}

We first consider the fully connected limit ${\alpha}=0$. The dynamics of $M_z(t)$ is shown in Fig.~\ref{fig:Long-range_entang}. Panels (a) and (b) compare different values of the dissipation strength $\chi$ and different system sizes.

For \(\chi=0.1\), shown in Fig.~\ref{fig:Long-range_entang}(a), the
magnetization exhibits robust oscillations whose amplitude becomes increasingly
stable as \(L\) grows. This behavior is consistent with the boundary-time-crystal
phase, where finite-size damping is suppressed in the thermodynamic limit. For
the larger value \(\chi=0.5\), shown in Fig.~\ref{fig:Long-range_entang}(b),
oscillations remain visible, but the finite-size damping is stronger. This is
expected because the system is closer to the transition at \(\chi=1\).

The Fourier spectrum in Fig.~\ref{fig:Long-range_entang}(c) shows a sharp
dominant peak in the oscillatory phase. The position of this peak is essentially
independent of system size, confirming that the observed oscillations correspond
to a well-defined collective frequency rather than to a finite-size transient.
The inset shows the dominant frequency as a function of \(L\) for different
values of \(\chi\). As \(\chi\) approaches and crosses the transition point, the
oscillation frequency decreases and eventually vanishes in the stationary
magnetized phase.

\begin{figure*}
    \centering
    \includegraphics[width=0.7\columnwidth]{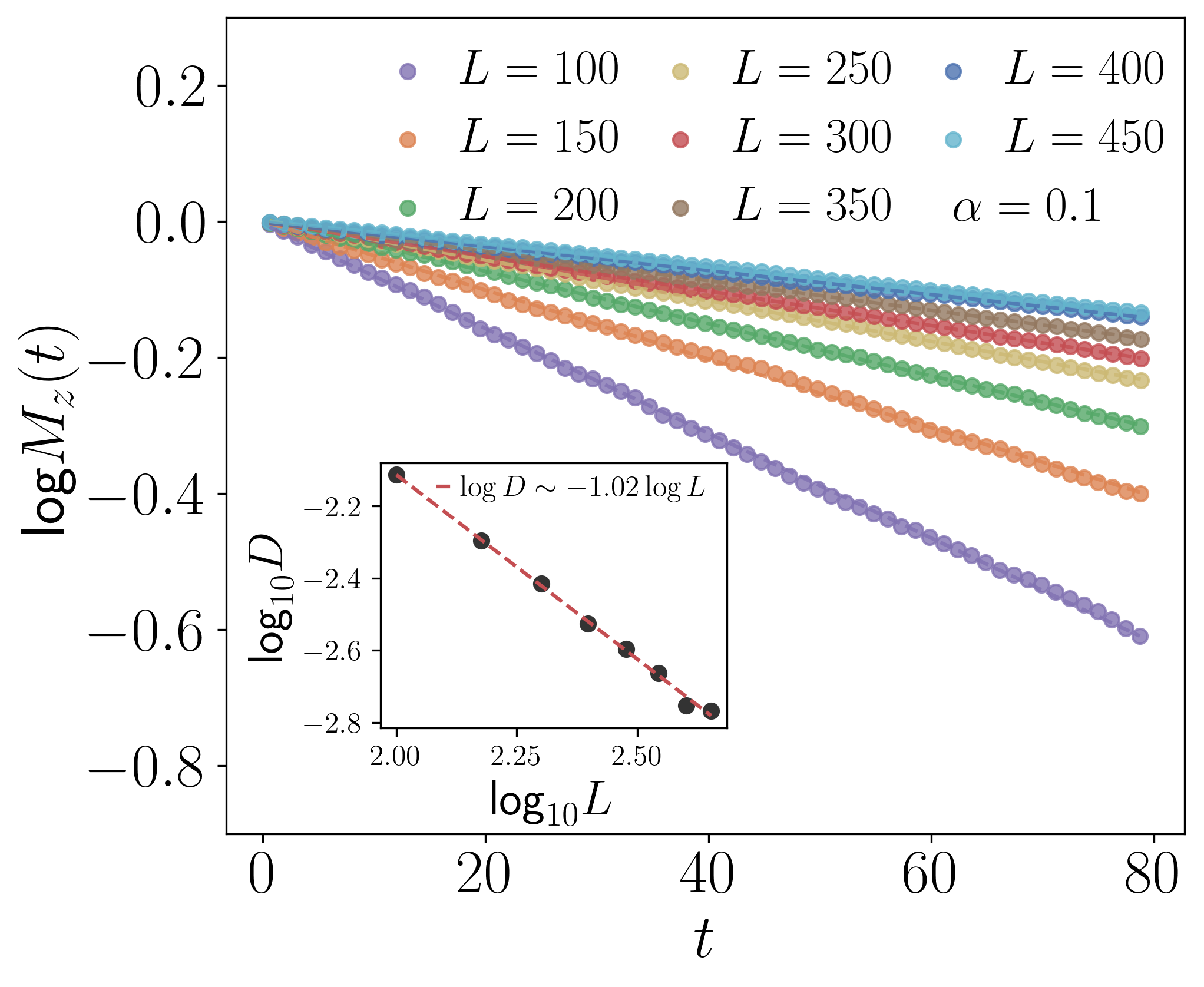}\put(-150,150){(a)}
    \includegraphics[width=0.7\columnwidth]{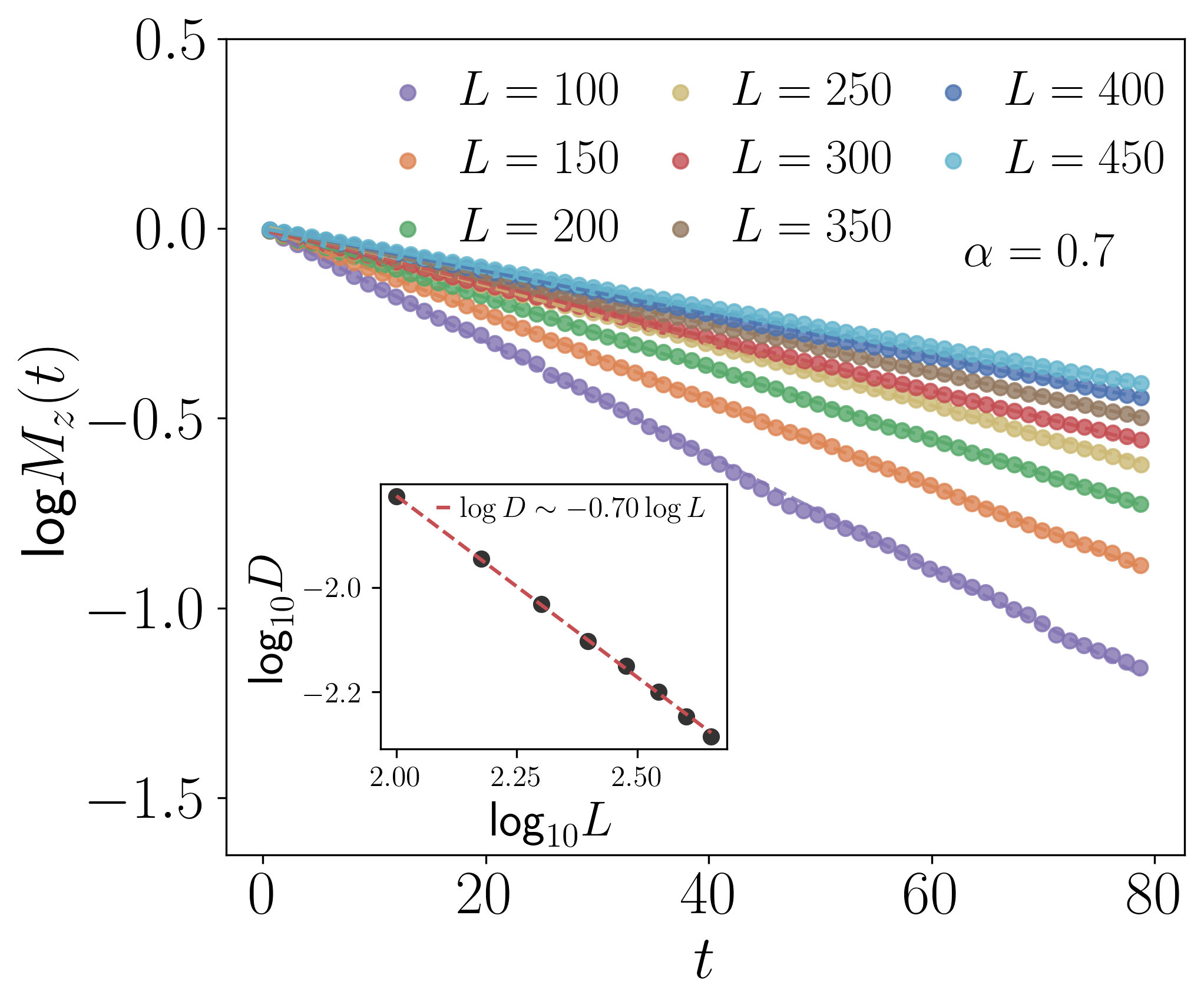}\put(-150,150){(b)}
    \includegraphics[width=0.7\columnwidth]{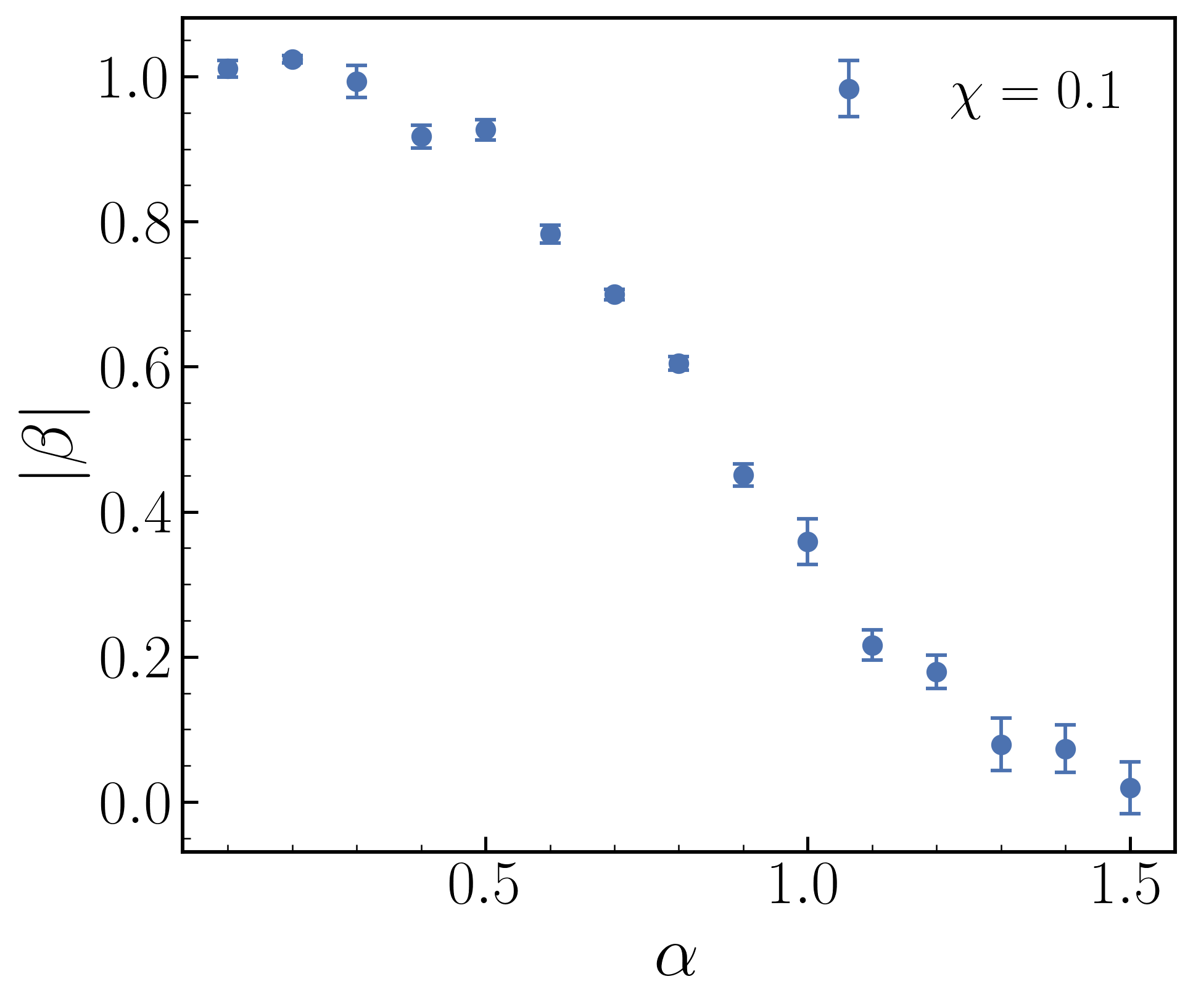}\put(-155,150){(c)}
    \caption{\textbf{Finite-size decay for nonzero power-law exponent ${\alpha}$.}
    (a),(b) Logarithm of the oscillation-envelope maxima as a function of time for different system sizes. The approximately linear behavior indicates an exponential finite-size decay of the envelope. Insets: decay rate $D$ as a function of $L$ in a log-log scale. The fits show $D(L)\sim L^{-\beta}$.
    (a) For ${\alpha}=0.1$, the decay exponent is approximately $\beta\simeq 1.02$.
    (b) For ${\alpha}=0.7$, the exponent is approximately $\beta\simeq 0.70$.
    (c) Scaling exponent $\beta$ as a function of ${\alpha}$. The exponent decreases with increasing ${\alpha}$ and becomes compatible with zero for ${\alpha}\gtrsim 1.2$, indicating the loss of persistent oscillations in the thermodynamic limit.}
    \label{fig:Long-range_entang1}
\end{figure*}

\subsection{Decay rate}

We next study the finite-size decay of the oscillations for nonzero power-law exponent ${\alpha}$. 
Because $M_z(t)$ oscillates approximately symmetrically around zero, the damping is extracted 
from the envelope rather than from the raw signal. We therefore identify the successive positive 
maxima $M_z^{\rm peak}(t_n)$ and fit their decay as
\begin{equation}
   M_z^{\rm peak}(t_n) \simeq A e^{-D t_n},
\end{equation}
where \(A\) is a nonuniversal amplitude and \(D\) is the finite-size decay rate. The dependence of this decay rate on system size is then fitted as
\begin{equation}
    D(L) \sim L^{-\beta}.
\end{equation}
Equivalently,
\begin{equation}
    \ln D = -\beta \ln L + \mathrm{const.}
\end{equation}
A positive value of \(\beta\) therefore means that the decay rate vanishes as
\(L\) increases, implying persistent oscillations in the thermodynamic limit.

The results are shown in Fig.~\ref{fig:Long-range_entang1}. For \({\alpha}=0.1\), Fig.~\ref{fig:Long-range_entang1}(a), the oscillation envelope
decays exponentially at finite size, but the decay becomes systematically slower
as \(L\) increases. The inset shows that the extracted decay rate follows a
power law with exponent approximately \(\beta\simeq 1.02\). This indicates a
rapid suppression of damping with system size.

For \({\alpha}=0.7\), Fig.~\ref{fig:Long-range_entang1}(b), the same qualitative
behavior is observed, but the scaling exponent is smaller, approximately
\(\beta\simeq 0.70\). Thus, the lifetime of the oscillations still grows with
system size, but more slowly than for smaller \({\alpha}\).

Finally, Fig.~\ref{fig:Long-range_entang1}(c) shows the exponent $\beta$ as a function of ${\alpha}$. The exponent decreases monotonically as the dissipative coupling becomes shorter ranged and approaches zero for ${\alpha} \gtrsim 1.2$. This suggests that long-range dissipation stabilizes the boundary-time-crystal oscillations by causing the finite-size decay rate to vanish algebraically with system size. Once $\beta$ becomes compatible with zero, the decay rate no longer decreases with $L$, and persistent oscillations are lost. Within our approximation, therefore, the time-crystalline regime extends beyond ${\alpha}=1$, with $\beta$ remaining positive over a finite range of ${\alpha}>1$. This is particularly noteworthy in the light of Ref.~\cite{Passarelli2022LongRangeLindbladians}, which established a time crystal for ${\alpha}<1$, where mean-field theory becomes exact in the thermodynamic limit, but left the fate of the system for ${\alpha}>1$ unresolved.

Our semiclassical stochastic results are therefore consistent with the
thermodynamic picture of the long-range Lindbladian model: sufficiently
long-ranged dissipation supports a boundary time crystal, while shorter-ranged
dissipation leads to relaxation. In addition, the present analysis provides a
direct finite-size dynamical characterization of the time-crystal lifetime
through the exponent \(\beta\).



\section{Spin-one model}\label{spin-one:sec}
\subsection{Hamiltonian and dissipators}

We now consider the spin-1 model of Ref.~\cite{Wang2025BoundaryTimeCrystals}. This model is conceptually different from the spin-$1/2$ long-range Lindbladian studied above. In the spin-$1/2$ case, the long-range character is encoded directly in the dissipative jump operators. Here, instead, the dissipation is strictly local, while the long-range coupling appears in the coherent Hamiltonian interaction. Therefore, this model provides a useful test of whether the Langevin approach can also capture a boundary time crystal generated by local decay together with long-range interactions.

The Hamiltonian is given by
\begin{equation}\label{eq:spinone_hamiltonian}
\hat{H} =
\frac{\Omega}{\sqrt{2}}\sum_{i=1}^L \hat{s}_i^x
-\Delta\sum_{i=1}^L \hat{n}_i
-E\sum_{i=1}^L \hat{s}_i^z
-\sum_{i<j}V_{ij}\hat{n}_i\hat{n}_j ,
\end{equation}
where the onsite operators are spin-1 ones, and one defines $\hat{n}_i=(\hat{s}_i^z)^2$. We have
\begin{equation}
  V_{ij} = \frac{C}{[D(|i-j|)]^\alpha}\,,
\end{equation}
with
\begin{equation}
C = \dfrac{\chi}{\sum_{r=1}^{L-1} [D(r)]^{-\alpha}}\,,
\end{equation}
and $\chi$ is a real positive parameter. For periodic boundary conditions, \(|i-j|\) is replaced by the shortest distance
on the ring. The exponent $\alpha$ controls the range of the coherent interaction. The case $\alpha=0$ corresponds to fully connected interactions, while increasing $\alpha$ makes the interaction progressively more short ranged. For $\alpha=0$ we take $C=\chi/L$, as in Ref.~\cite{Wang2025BoundaryTimeCrystals}. For $\alpha>0$, we use the Kac-normalized coupling above in order to keep the interaction energy per spin finite as the system size is increased.

In the Lindbladian there are $2L$ Lindblad operators defined as
\begin{equation}\label{lops1:eqn}
  \hat{L}_{j\,\pm} = \ket{0}_j\bra{\pm}\,,
\end{equation}
where $\ket{0}_j$ and $\ket{\pm}_j$ are the eigenstates of $\hat{s}_j^z$. These jump operators describe independent local decay from the excited states $\ket{\pm}$ to the state $\ket{0}$. Thus, in contrast to collective-dissipation BTC models, no nonlocal dissipative channel is imposed. Any collective oscillation must therefore emerge from the competition between local dissipation and long-range Hamiltonian interactions.

Since the local Hilbert space is three-dimensional, we formulate the
semiclassical dynamics in a basis of Gell-Mann matrices, as explained in Ref.~\cite{Zhu_2019}. This provides a
complete set of local traceless operators for the spin-one problem and leads to
the stochastic equations derived in the next subsections.

\subsection{Gell-Mann representation}
We use the following normalized Gell-Mann basis for the local three-dimensional
Hilbert space. In the ordered basis \(\{\ket{+},\ket{0},\ket{-}\}\), the
matrices are
%
\begin{align}
\hat{\Lambda}_1 &=
\frac{1}{\sqrt{2}}
\begin{pmatrix}
0&1&0\\
1&0&0\\
0&0&0
\end{pmatrix},
&
\hat{\Lambda}_2 &=
\frac{1}{\sqrt{2}}
\begin{pmatrix}
0&0&1\\
0&0&0\\
1&0&0
\end{pmatrix},
\nonumber\\[0.8em]
\hat{\Lambda}_3 &=
\frac{1}{\sqrt{2}}
\begin{pmatrix}
0&0&0\\
0&0&1\\
0&1&0
\end{pmatrix},
&
\hat{\Lambda}_4 &=
\frac{1}{\sqrt{2}}
\begin{pmatrix}
0&-i&0\\
i&0&0\\
0&0&0
\end{pmatrix},
\nonumber\\[0.8em]
\hat{\Lambda}_5 &=
\frac{1}{\sqrt{2}}
\begin{pmatrix}
0&0&-i\\
0&0&0\\
i&0&0
\end{pmatrix},
&
\hat{\Lambda}_6 &=
\frac{1}{\sqrt{2}}
\begin{pmatrix}
0&0&0\\
0&0&-i\\
0&i&0
\end{pmatrix},
\nonumber\\[0.8em]
\hat{\Lambda}_7 &=
\frac{1}{\sqrt{2}}
\begin{pmatrix}
1&0&0\\
0&-1&0\\
0&0&0
\end{pmatrix},
&
\hat{\Lambda}_8 &=
\frac{1}{\sqrt{6}}
\begin{pmatrix}
1&0&0\\
0&1&0\\
0&0&-2
\end{pmatrix}.
\end{align}
%
They are normalized such that
\[
\mathrm{Tr}(\hat{\Lambda}_\mu\hat{\Lambda}_\nu)=\delta_{\mu\nu}.
\]
In this basis, the spin-one angular momentum operators are represented as
\begin{align}
\hat{S}_x &= \hat{\Lambda}_1 + \hat{\Lambda}_3, \\
\hat{S}_y &= \hat{\Lambda}_4 + \hat{\Lambda}_6, \\
\hat{S}_z &= \frac{1}{\sqrt{2}}\,\hat{\Lambda}_7 + \frac{\sqrt{3}}{\sqrt{2}}\,\hat{\Lambda}_8.
\end{align}
Since the Hamiltonian contains
\(\hat{n}_i=(\hat{S}_i^z)^2\), we also need its representation in the same basis:
\begin{equation}\label{nino:eqn}
 \hat{n}_i = \frac{1}{4}\left(\sqrt{2}\hat{\Lambda}_{7\,i}+\sqrt{6}\hat{\Lambda}_{8\,i}\right)^2 =\frac23\boldsymbol{1}-\frac{\sqrt{6}}{6}\hat{\Lambda}_{8\,i}+\frac{\sqrt{2}}{2}\hat{\Lambda}_{7\,i}\,.
\end{equation}
The Gell-Mann representation is useful because the spin-1 Hilbert space is three-dimensional, and the eight matrices $\hat{\Lambda}_{\mu}$ form a complete basis for the traceless local operators. Using the relations above, the Hamiltonian in Eq.~\eqref{eq:spinone_hamiltonian}
can be rewritten as
\begin{equation}
\label{eq:spinone_hamiltonian_gellmann}
\begin{split}
\hat H
={}&
\sum_{i=1}^L
\bigg[
\frac{\Omega}{\sqrt{2}}
\left(
\hat{\Lambda}_{1,i}
+
\hat{\Lambda}_{3,i}
\right)
+
\frac{\Delta-3E}{\sqrt{6}}
\hat{\Lambda}_{8,i}
\\
&\hspace{1.1cm}
-
\frac{\Delta+E}{\sqrt{2}}
\hat{\Lambda}_{7,i}
-
\frac{2\Delta}{3}\boldsymbol{1}
\bigg]
-
\sum_{i<j}
V_{ij}\hat n_i\hat n_j .
\end{split}
\end{equation}
%
In the Gell-Mann basis the Lindblad operators Eq.~\eqref{lops1:eqn} operators read
\begin{align}
\label{eq:spinone_jump_gellmann}
  \hat{{L}}_{\,+j}
  &=
  \frac{1}{\sqrt{2}}
  \left(
  \hat{\Lambda}_{1j}
  -
  i\hat{\Lambda}_{4j}
  \right),
  \nonumber\\
  \hat{{L}}_{j\,-}
  &=
  \frac{1}{\sqrt{2}}
  \left(
  \hat{\Lambda}_{3j}
  +
  i\hat{\Lambda}_{6j}
  \right).
\end{align}
These relations express the local decay processes
\(\ket{\pm}_j\to\ket{0}_j\) in the same operator basis used for the
semiclassical variables.
\subsection{Semiclassical Langevin equations in the Gell-Mann basis}
\label{subsec:spinone_semiclassical}

Having expressed both the Hamiltonian and the local jump operators in the
Gell-Mann basis, we now introduce the corresponding semiclassical variables
\begin{equation}
  \lambda_{\mu j}(t)=\langle \hat{\Lambda}_{\mu j}(t)\rangle,
  \qquad \mu=1,\ldots,8 .
\end{equation}
The local magnetization is reconstructed as
\begin{equation}
  s_j^z(t)
  =
  \frac{1}{\sqrt{2}}\lambda_{7j}(t)
  +
  \frac{\sqrt{3}}{\sqrt{2}}\lambda_{8j}(t).
\end{equation}
The longitudinal magnetization is then obtained from these local variables
according to Eq.~\eqref{eq:Mz_traj_avg}.

Starting from the quantum Langevin equation and factorizing operator products at
the semiclassical level, the equations of motion can be written in the compact
form
\begin{equation}
\label{eq:spinone_sde_compact}
  \dot{\lambda}_{\mu j}
  =
  A_{\mu j}(\boldsymbol{\lambda})
  +
  \sqrt{\gamma}\,
  D_{\mu j}(\boldsymbol{\lambda},\boldsymbol{\xi}),
  \qquad
  \mu=1,\ldots,8 .
\end{equation}
Here \(A_{\mu j}\) contains the deterministic coherent and dissipative drift
terms, while \(D_{\mu j}\) contains the multiplicative noise generated by the
local decay channels. Their explicit expressions are given in
Appendix~\ref{app:spinone_langevin_derivation}. $\boldsymbol{\lambda}$ marks the set of the $\lambda_\mu(t)$, while $\boldsymbol{\xi}$ marks the set of the real Gaussian noises that satisfy
\begin{equation}
  \left\langle
  \xi_{\sigma j}^{(a)}(t)
  \xi_{\tau l}^{(b)}(t')
  \right\rangle
  =
  \delta^{ab}\delta_{\sigma\tau}\delta_{jl}\delta(t-t') .
\end{equation}
Equation~\eqref{eq:spinone_sde_compact} retains the deterministic nonlinear
mean-field dynamics together with the stochastic corrections generated by the
local dissipative channels.
The derivation of Eq.~\eqref{eq:spinone_sde_compact} from the operator Langevin equations is given in Appendix~\ref{app:spinone_langevin_derivation}. These equations retain the deterministic nonlinear dynamics together with the leading stochastic corrections generated by the local dissipative channels.

For \(\alpha=0\), the coherent interaction is fully connected. In this limit,
the collective Gell-Mann variables become classical in the thermodynamic limit.
Indeed, defining
\(\hat{\Lambda}_{\mu,c}=\sum_j\hat{\Lambda}_{\mu j}\) and
\(\hat{\lambda}_\mu=\hat{\Lambda}_{\mu,c}/L\), their commutators scale as
\(1/L\) and vanish for \(L\to\infty\). The dynamics therefore becomes
mean-field-like in the thermodynamic limit, while
Eq.~\eqref{eq:spinone_sde_compact} describes finite-size stochastic corrections
around this limit.


\section{Numerical results for the spin-one model}\label{numerical1:sec}
To study the dynamics, we solve the coupled stochastic equations for the
Gell-Mann variables of the spin-one model, written compactly in
Eq.~\eqref{eq:spinone_sde_compact}. We focus on the trajectory-averaged
magnetization \(M_z(t)\), defined in Eq.~\eqref{eq:Mz_traj_avg}, and on its
dependence on system size. This allows us to determine whether the oscillatory
dynamics becomes increasingly stable as the coherent interaction becomes more
collective.

We use a classical initialization in which all spins are prepared in the
\(\hat S_j^z\) eigenstate with \(\langle S_j^z\rangle=1\). In the Gell-Mann
variables this corresponds to
\[
\lambda_{7j}=\frac{1}{\sqrt{2}},
\qquad
\lambda_{8j}=\frac{1}{\sqrt{6}},
\]
with all other \(\lambda_{\mu j}\) initially set to zero. We also tested a discrete truncated-Wigner initialization of the spin-one
variables, following Ref.~\cite{Zhu_2019}. This did not appreciably affect the
magnetization dynamics or the extracted finite-size scaling exponents, so the
results shown below are obtained using the classical initialization.
\subsection{Fully connected dynamics}
\begin{figure}[H]
    \centering
    \includegraphics[width=0.7\columnwidth]{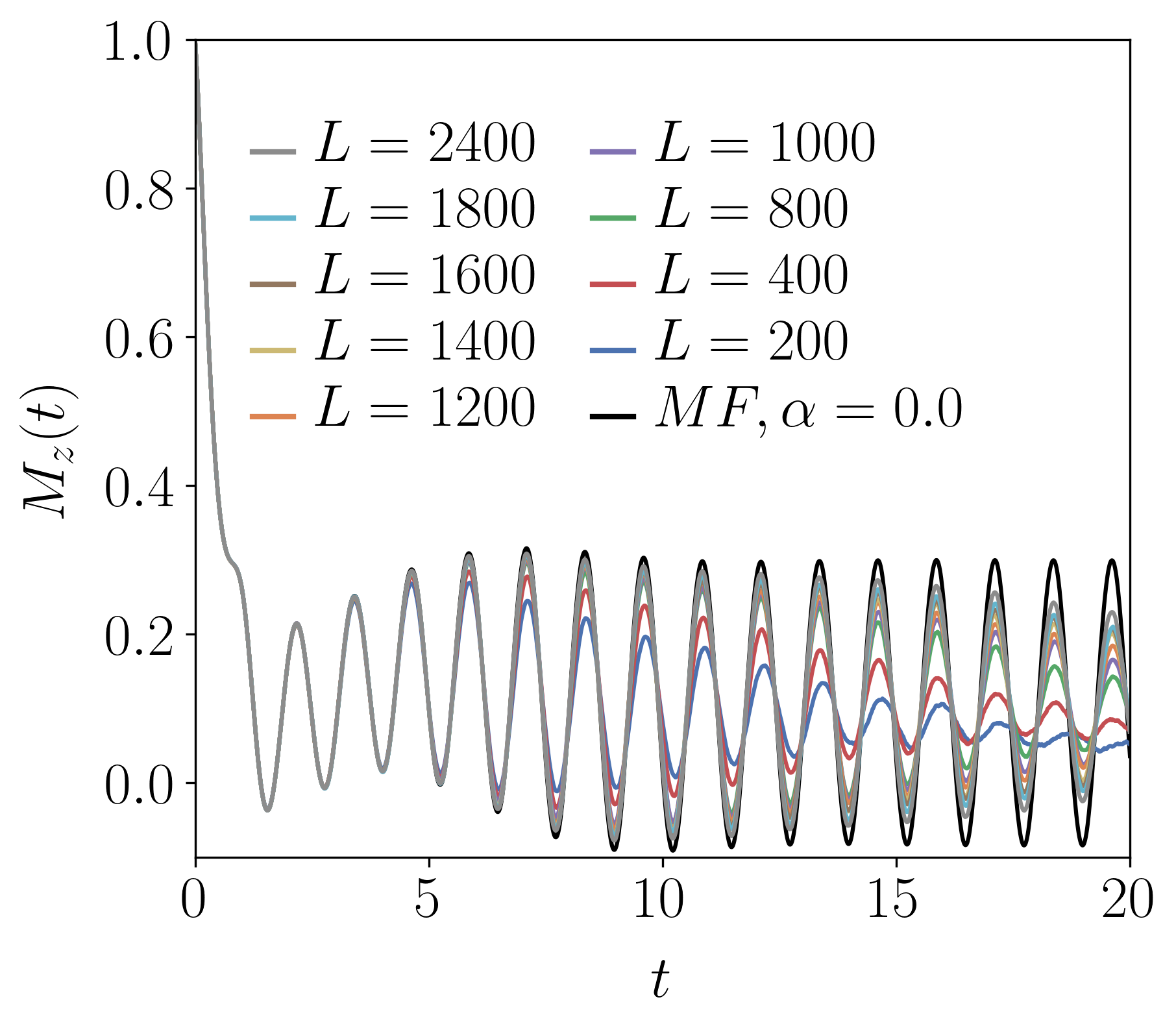}\put(-150,152){(a)}\\
    \includegraphics[width=0.7\columnwidth]{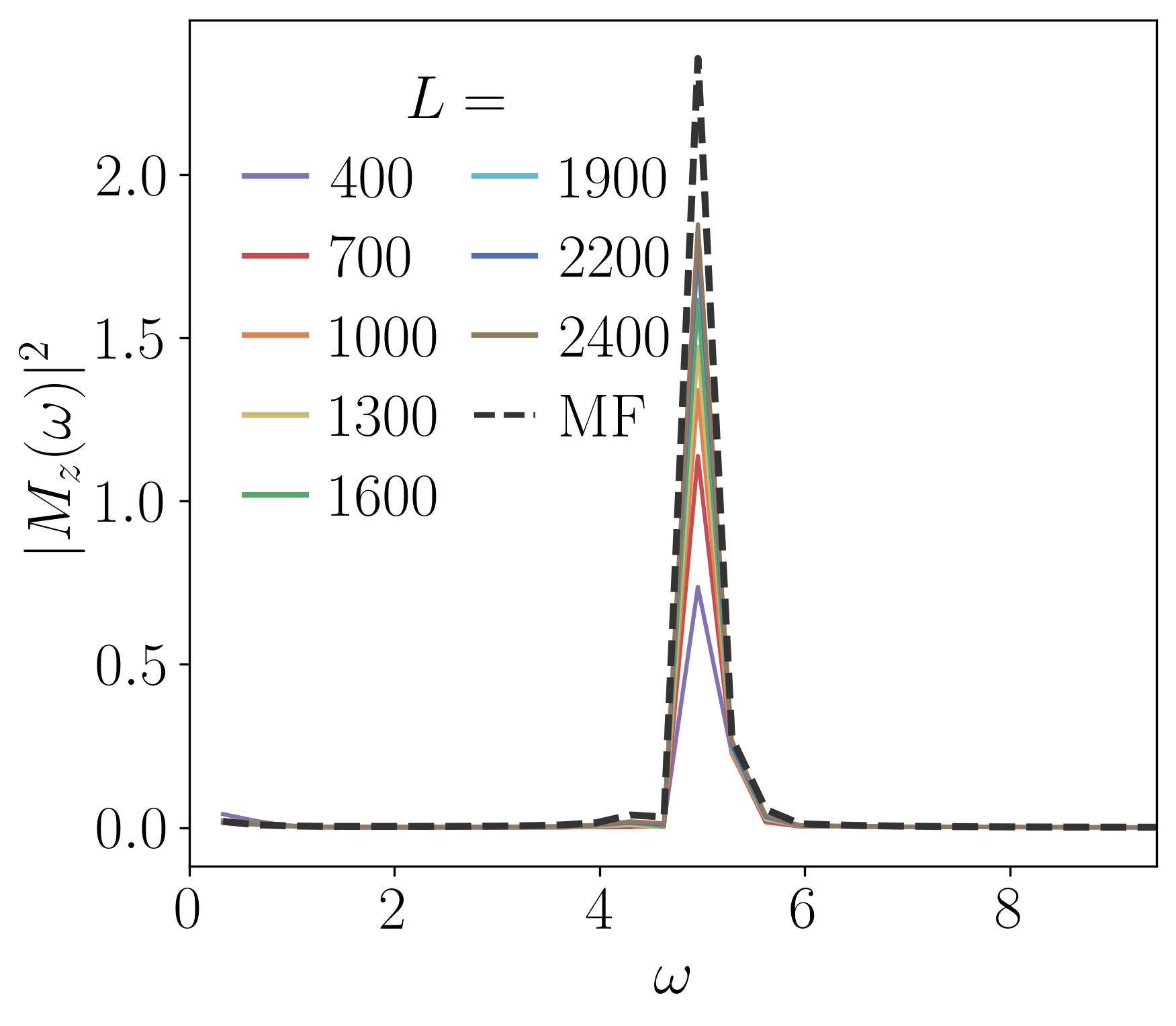}\put(-150,152){(b)}
    \caption{
\textbf{Spin-one model with fully connected interactions.}
(a) Time evolution of \(M_z(t)\) obtained from the semiclassical Langevin
equations, Eq.~\eqref{eq:spinone_sde_compact}, for an initially polarized state
with \(\langle S_z\rangle=1\), corresponding to
\(\lambda_7=1/\sqrt{2}\), \(\lambda_8=1/\sqrt{6}\), and all other
\(\lambda_\mu\) initially set to zero. The black curve denotes the mean-field
result. As \(L\) increases, the trajectory-averaged stochastic dynamics
approaches the mean-field limit cycle.
(b) Fourier spectrum of \(M_z(t)\). A dominant peak appears at the collective
oscillation frequency. The peak position agrees with the mean-field frequency,
while the spectral response approaches the mean-field result as \(L\) increases.
Model parameters: \(\Omega=4.0\), \(\Delta=-8.9\), \(E=4.0\), and \(\chi=16\).
}
    \label{fig:spinone_a0}
\end{figure}

We first consider the stochastic dynamics in the fully connected case
\(\alpha=0\), where the coherent interaction is collective. The results are
shown in Fig.~\ref{fig:spinone_a0}. As the system size is increased, the
trajectory-averaged dynamics approaches the mean-field solution. In the
large-system limit, the stochastic corrections become negligible and the
dynamics becomes indistinguishable from the mean-field prediction, confirming
that finite-size fluctuations are suppressed as \(L\) grows. The Fourier
spectrum in Fig.~\ref{fig:spinone_a0}(b) shows a pronounced peak near the
mean-field frequency, with the spectral response converging toward the
mean-field result as the system size increases.

\subsection{Finite-size scaling of the deviation time}
\begin{figure}
    \centering
    \includegraphics[width=70mm]{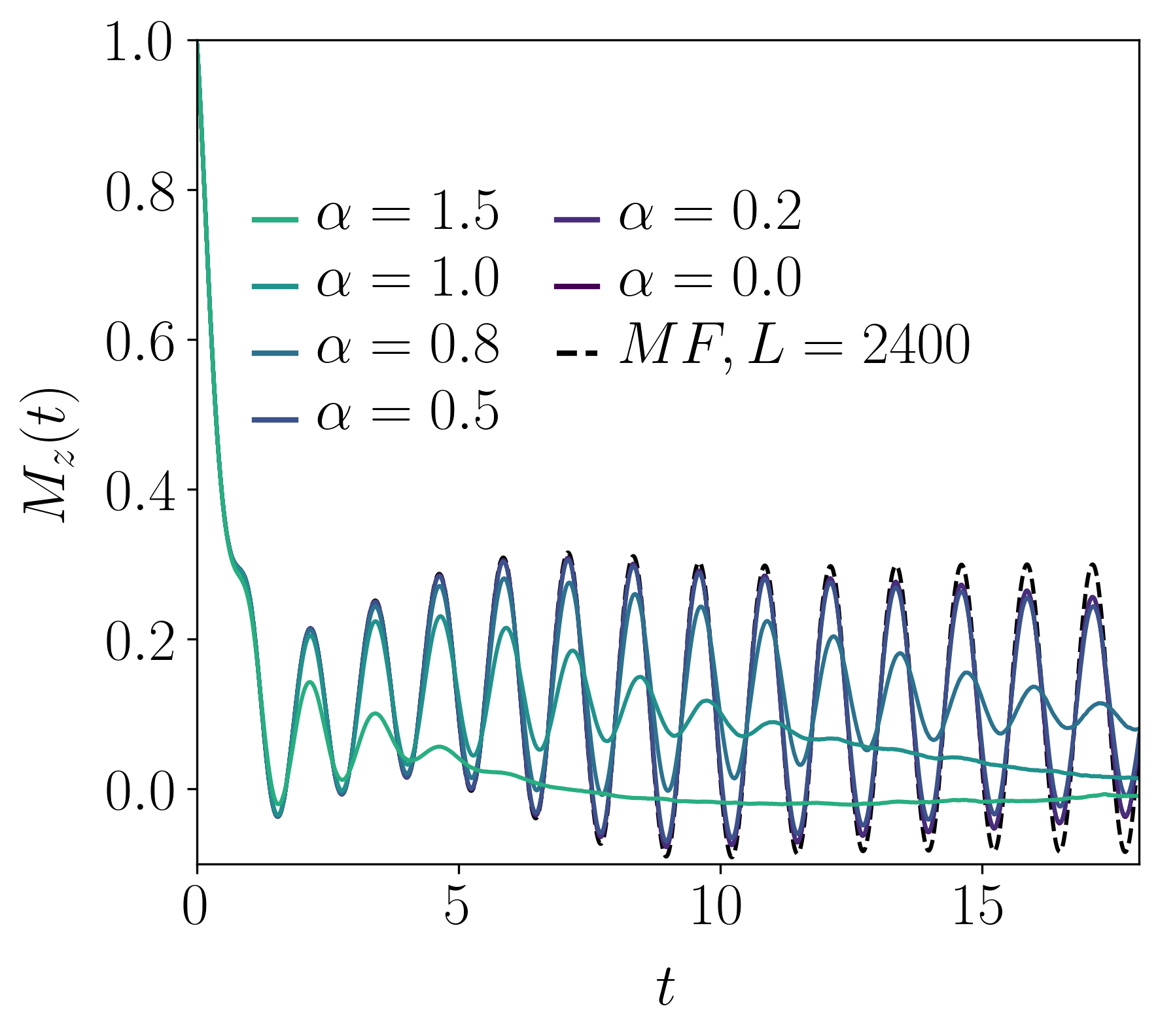}
    \caption{\textbf{Spin-one model with fully-finite-$\alpha$ interaction}.  Here, we compare the dynamics of $M_z$ for different values of $\alpha$. For $\alpha \leq 0.5$, the dynamics remains close to the mean-field prediction. At $\alpha=0.8$, we observe an intermediate level of damping. For $\alpha \geq 1$, the damping becomes strong, causing the oscillations to decay rapidly at early times.
. Numerical parameters: $\Omega = 3.0$, $\Delta = -7.0$, $E = 4.0$, $\chi = 1.0$.}
    \label{fig:damping_spin_one}
\end{figure}

{Let us now consider the persistence of the time-translation symmetry breaking oscillations for a generic value of $\alpha$. Before performing finite-size scalings,we illustrate the effect of the interaction range in Fig.~\ref{fig:damping_spin_one} where we compare the magnetization dynamics for several values of $\alpha$ at the largest system size we can attain, $L=2400$. In the strongly long-range regime, $\alpha\leq0.5$, the dynamics remains close to the fully connected mean-field trajectory and the oscillations exhibit only weak damping. For $\alpha=0.8$, deviations become more pronounced and an intermediate damping of the oscillations is observed. Upon increasing the exponent to $\alpha\geq1$, the oscillations decay rapidly already at early times, showing that finite-size fluctuations become increasingly important as the interaction becomes shorter ranged. So, in our semiclassical Langevin approximation, there seems to be no more a time-crystal behavior for $\alpha\geq 1$. This qualitative statement is confirmed by the finite-size scaling of the deviation time from the mean-field dynamics we are going to analyze in Sec.~\ref{devi:sec}. In Sec.~\ref{spec:sec} we study the finite-size scaling of the peak in the Fourier spectrum, and we find a power-law increase with the system size for $\alpha\leq 1.2$. Therefore the critical value of $\alpha$ beyond which there is no time crystal is between 1 and 1.5, in full agreement with~\cite{Wang2025BoundaryTimeCrystals}.}
\subsubsection{Finite-size scaling of the deviation time from the mean-field dynamics}\label{devi:sec}

When $\alpha \leq 1$ we expect that the thermodynamic-limit dynamics is described by the Lindblad mean-field one~\cite{Wang2025BoundaryTimeCrystals}, so it is meaningful to compare the finite-size Langevin mean-field results with that limit. For the spin-one model, the oscillations of $M_z(t)$ are not symmetric around zero. Therefore, instead of extracting a decay rate from the oscillation envelope, we quantify finite-size stability by comparing the stochastic finite-size signal directly with the mean-field trajectory. For each system size $L$, we define a deviation time $t^\star(L)$ as the time at which the finite-size dynamics departs persistently from the mean-field solution.

\begin{figure*}
    \centering

    \makebox[0.32\textwidth][c]{%
        \raisebox{0pt}{%
            \includegraphics[width=0.65\columnwidth]{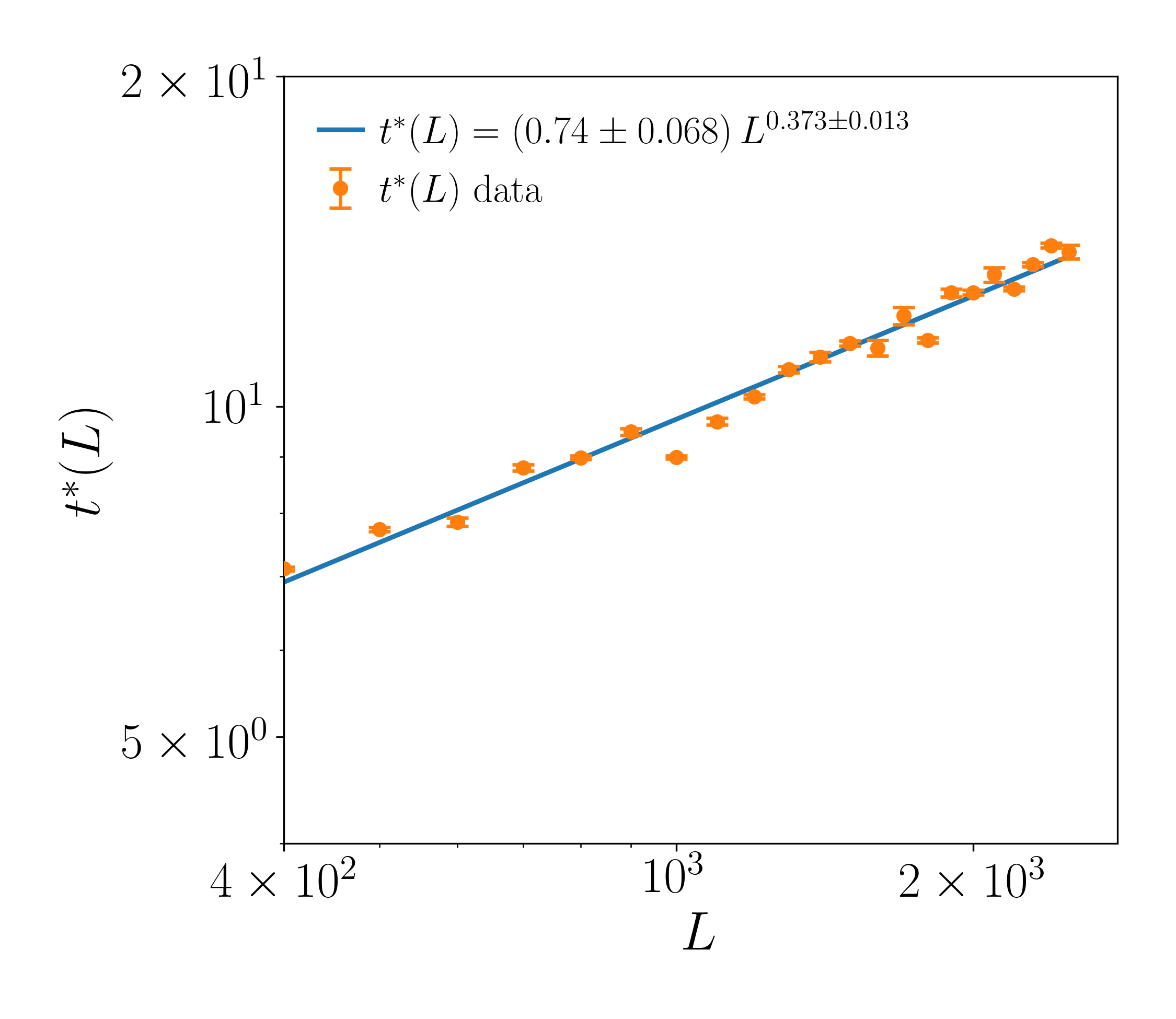}%
        }%
        \put(-150,135){(a)}%
        \put(-70,50){$\alpha=0.0$}%
    }
    \hfill
    \makebox[0.32\textwidth][c]{%
        \raisebox{0pt}{%
            \includegraphics[width=0.65\columnwidth]{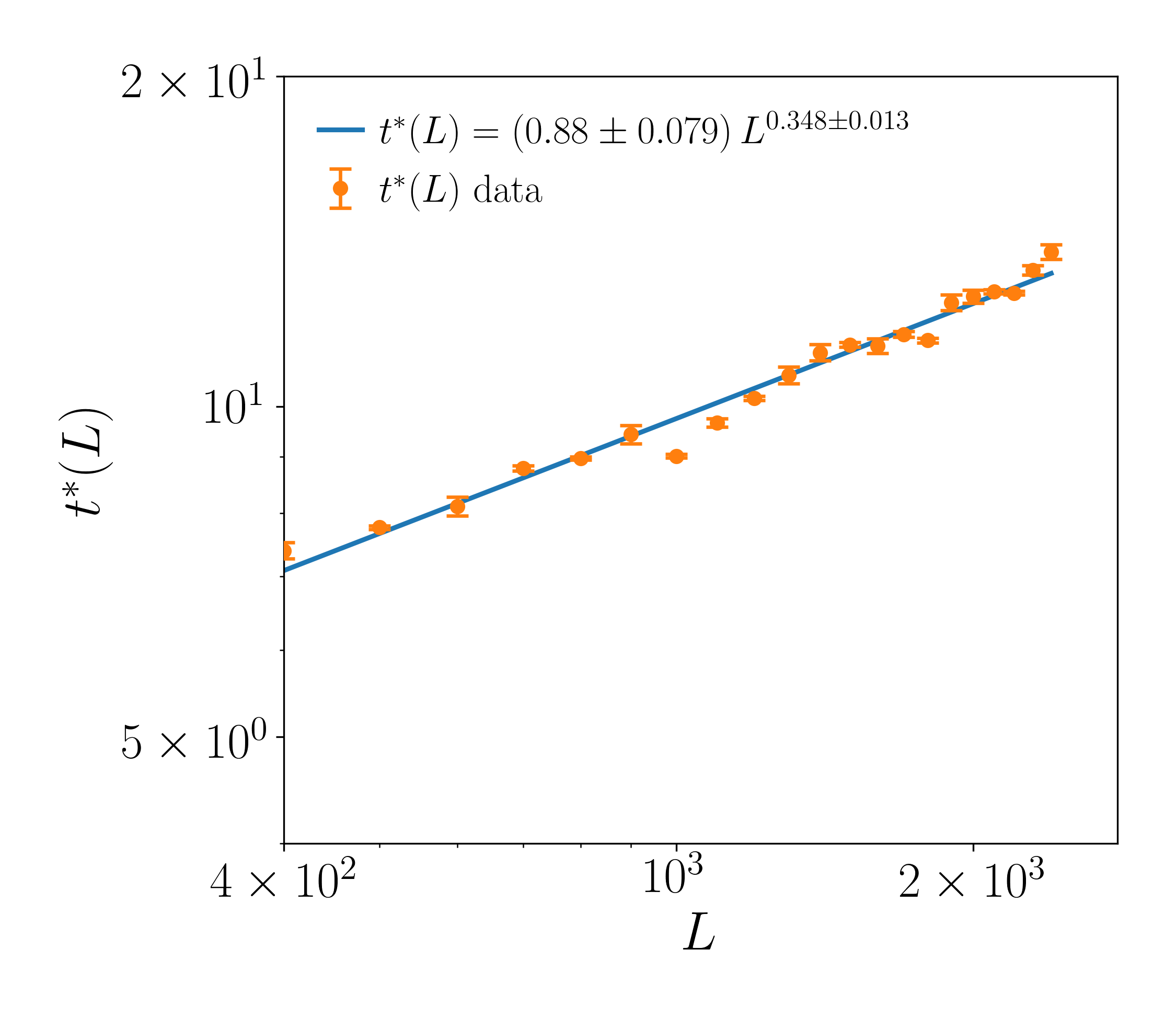}%
        }%
        \put(-150,135){(b)}%
        \put(-70,50){$\alpha=0.5$}%
    }
    \hfill
    \makebox[0.32\textwidth][c]{%
        \raisebox{8pt}{%
            \includegraphics[width=0.63\columnwidth]{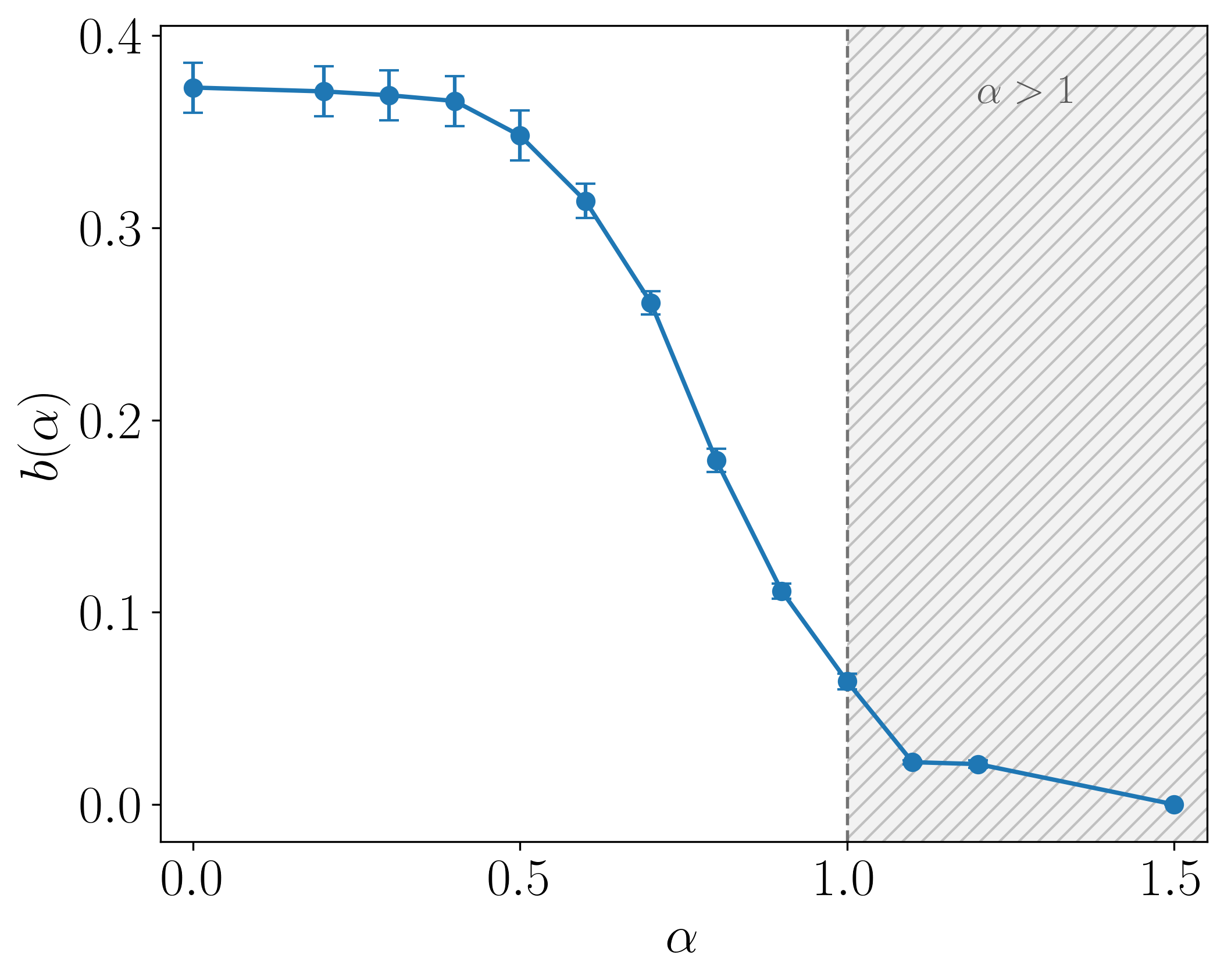}%
        }%
        \put(-150,135){(c)}%
    }

   \caption{
\textbf{Deviation time from mean-field dynamics in the spin-one model.}
(a),(b) Deviation time \(t^\star(L)\) as a function of system size for
\(\alpha=0\) and \(\alpha=0.5\). The data are fitted by
\(t^\star(L)=aL^b\). The extracted exponents are
\(b=0.37\pm0.013\) for \(\alpha=0\) and \(b=0.35\pm0.013\) for
\(\alpha=0.5\).
(c) Scaling exponent \(b(\alpha)\) as a function of the interaction exponent
\(\alpha\). The exponent remains finite in the long-range regime and decreases
as \(\alpha\) approaches one, indicating that the mean-field validity time no
longer grows appreciably with system size.
Model parameters: \(\Omega=4.0\), \(\Delta=-8.9\), \(E=4.0\), and \(\chi=16\).
}
    \label{fig:time_dev_MF}
\end{figure*}

To evaluate $t^*(L)$, and so quantify the finite-size stability of the spin-one oscillations, we compare
the finite-size trajectory-averaged signal \(M_z^{(L)}(t)\) with the mean-field
limit-cycle trajectory \(M_z^{\mathrm{MF}}(t)\). Since both signals are
oscillatory, an instantaneous distance would be sensitive to small phase shifts
and local fluctuations. We therefore use the accumulated squared deviation
\begin{equation}
\mathcal D_L(t)
=
\int_0^t
\left[
M_z^{(L)}(t')
-
M_z^{\mathrm{MF}}(t')
\right]^2
\,dt' .
\end{equation}
This quantity measures the total departure of the finite-size dynamics from the
mean-field oscillation up to time \(t\).

The deviation time \(t^\star(L)\) is then defined as the first time at which the
accumulated deviation reaches a fixed fraction of a reference value,
\begin{equation}
\mathcal D_L\!\left(t^\star\right)
=
q\mathcal D_{\mathrm{ref}} .
\end{equation}
Here \(\mathcal D_{\mathrm{ref}}\) is taken from the largest simulated system at
the final time of the evolution, and \(q\) is a dimensionless threshold
fraction. In the analysis we use \(q=0.12\). We have checked that the extracted
scaling exponent is stable under moderate changes of \(q\), so that the result
is not sensitive to the particular threshold choice.

We study the finite-size scaling of the deviation time by fitting
\begin{equation}
t^\star(L)=aL^b .
\end{equation}
A positive exponent \(b\) indicates that the time over which the finite-size
stochastic dynamics remains close to the mean-field limit cycle grows with
system size. Thus, \(b\) provides a finite-size measure of the stability of the
oscillatory phase.

Figure~\ref{fig:time_dev_MF}(c) shows the exponent \(b\) as a function of
\(\alpha\). For \(\alpha\lesssim0.5\), the exponent remains close to \(0.4\),
indicating robust algebraic growth of the mean-field validity time. As
\(\alpha\) approaches one, \(b\) decreases strongly and becomes close to zero,
signaling that the deviation time no longer grows significantly with system size
in the shorter-range regime.

Overall, the Langevin results are consistent with the physical picture of
Ref.~\cite{Wang2025BoundaryTimeCrystals}. Wang \emph{et al.} showed that local
dissipation can induce boundary-time-crystalline dynamics when combined with
sufficiently long-range coherent interactions. In their analysis, the
oscillation lifetime grows with system size in the long-range regime and ceases
to grow once the interactions become too short ranged.

Our results provide a complementary finite-size characterization of the same
mechanism. We find that the deviation time grows algebraically,
$t^\star(L)\sim L^b$, for small $\alpha$. In the fully connected case, for
example, we obtain $b\simeq 0.37$, showing that the finite-size stochastic
dynamics remains close to the mean-field oscillation for increasingly long
times as $L$ grows. As the interaction exponent is increased, $b(\alpha)$
decreases and becomes very small near $\alpha\simeq 1$, indicating that the
classical mean-field limit cycle is no longer efficiently stabilized by
increasing the system size. {We remark that these results are valid for $\alpha\leq 1$, because only in this range the mean-field dynamics describes the thermodynamic-limit behavior~\cite{Wang2025BoundaryTimeCrystals} and the accumulated squared deviation from this limit is meaningful. That's why in Fig. 5(d) we shade the region $\alpha \geq 1$. In this region applying our method provides a deviation time scaling very slowly with the system size, for $\alpha \leq 1.2$. This is in agreement with the results of~\cite{Wang2025BoundaryTimeCrystals} but it is most probably an artifact of our method that, like all semiclassical methods, works fine only when the system is enough long range.}

The deviation time $t^\star(L)$ may also be interpreted as an Ehrenfest time,
namely, the time scale beyond which finite-size quantum fluctuations cause the
quantum dynamics to depart appreciably from the classical mean-field
trajectory. Its algebraic scaling with $L$, rather than a logarithmic scaling,
suggests that initially nearby trajectories, separated on a scale set by the
effective Planck constant $1/N$, diverge only algebraically in time. This
behavior is therefore indicative of regular, rather than chaotic,
dynamics~\cite{Pappalardi_2018}.

{In
Ref.~\cite{Wang2025BoundaryTimeCrystals} the distinction was made between the classical long-range BTC
regime and the more strongly correlated regime at larger $\alpha$: Our
semiclassical Langevin approach directly probes the former.} In this regime it shows how the
persistent mean-field oscillation is recovered from finite-size stochastic
dynamics and provides the exponent $b(\alpha)$ as a quantitative measure of
the robustness of this classical time-crystalline regime. Interestingly,
$b(\alpha)$ remains approximately constant only over the interval
$\alpha\in[0,0.4]$, {and then decreases monotonically, attaining a value much smaller than 1 for $\alpha = 1$}. Thus, the range in which the finite-size dynamics displays
a robust mean-field-like scaling is narrower than the interval
$\alpha\in[0,1]$ over which mean-field theory is valid in the thermodynamic
limit. {So, although for $\alpha\leq 1$ the thermodynamic limit is mean-field, the finite-size scaling towards this limit is not.}
\subsubsection{Finite-size scaling of the dominant Fourier peak}
\label{spec:sec}
We investigate the finite-size scaling of the dominant peak in the
Fourier spectrum of the magnetization. A peak increasing as a power law with the system size is a mark of time-translation symmetry breaking oscillations that last more and more for increasing system size, that's to say a mark of a time-crystal behavior. In contrast with the analysis of the deviation time considered above, here there is no reference to a thermodynamic-limit behavior, and so results obtained with this method are valid also beyond the range $\alpha\in[0,1]$ where the thermodynamic-limit behavior is described by the mean-field dynamics. 

For a system of size $L$, we define
the Fourier transform over the fixed time interval $t_{\min}\leq t\leq
t_{\max}$ as
\begin{equation}
M_z(\omega,L)
=
\Delta t
\sum_{n}
M_z(t_n,L)\,
e^{-i\omega t_n},
\label{eq:Mz_Fourier}
\end{equation}
where $\Delta t$ is the time step and the sum is restricted to
$t_{\min}\leq t_n\leq t_{\max}$. In the numerical analysis, the same time
window, $1\leq t\leq20$, is used for all system sizes and for the
mean-field result.

We characterize the height of the dominant Fourier peak through its power,
\begin{equation}
\mathcal{P}(L)
\equiv
\max_{\omega>0}
\left|M_z(\omega,L)\right|^2,
\label{eq:peak_power_definition}
\end{equation}
where the dependence on $\alpha$ is left implicit. Notice that
$\mathcal{P}(L)$ denotes the peak of the Fourier power spectrum, rather than
the peak amplitude itself.

To determine the finite-size behavior, we assume an algebraic scaling form
\begin{equation}
\mathcal{P}(L)
=
C(\alpha)L^{s(\alpha)},
\label{eq:peak_power_scaling}
\end{equation}
where $C(\alpha)$ is a size-independent prefactor and $s(\alpha)$ is the
finite-size scaling exponent. Taking the logarithm of
Eq.~\eqref{eq:peak_power_scaling} gives
\begin{equation}
\log \mathcal{P}(L)
=
b(\alpha)
+
s(\alpha)\log L,
\qquad
b(\alpha)=\log C(\alpha).
\label{eq:peak_power_log_scaling}
\end{equation}
We therefore extract $s(\alpha)$ directly from a linear least-squares fit
of $\log \mathcal{P}(L)$ as a function of $\log L$. No nonlinear
thermodynamic-limit extrapolation is performed.

\begin{figure}
    \centering
\begin{tabular}{c}
    \makebox[0.4\textwidth][c]{%
        \raisebox{0pt}{%
            \includegraphics[width=0.7\columnwidth]{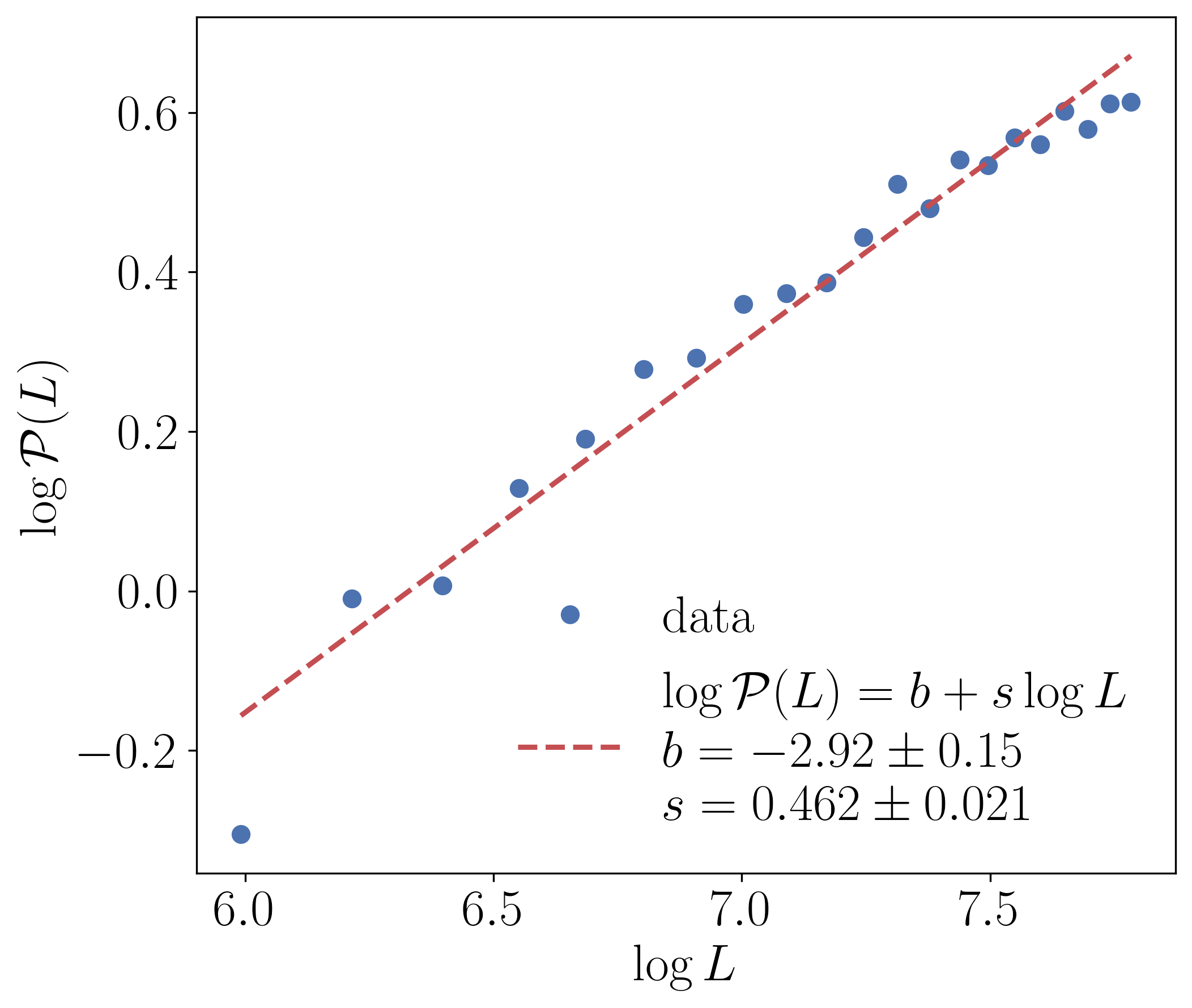}%
        }%
        \put(-165,143){(a)}%
        \put(-95,130){$\alpha=0.0$}%
    }\\
    \hfill
    \makebox[0.4\textwidth][c]{%
        \raisebox{0pt}{%
            \includegraphics[width=0.7\columnwidth]{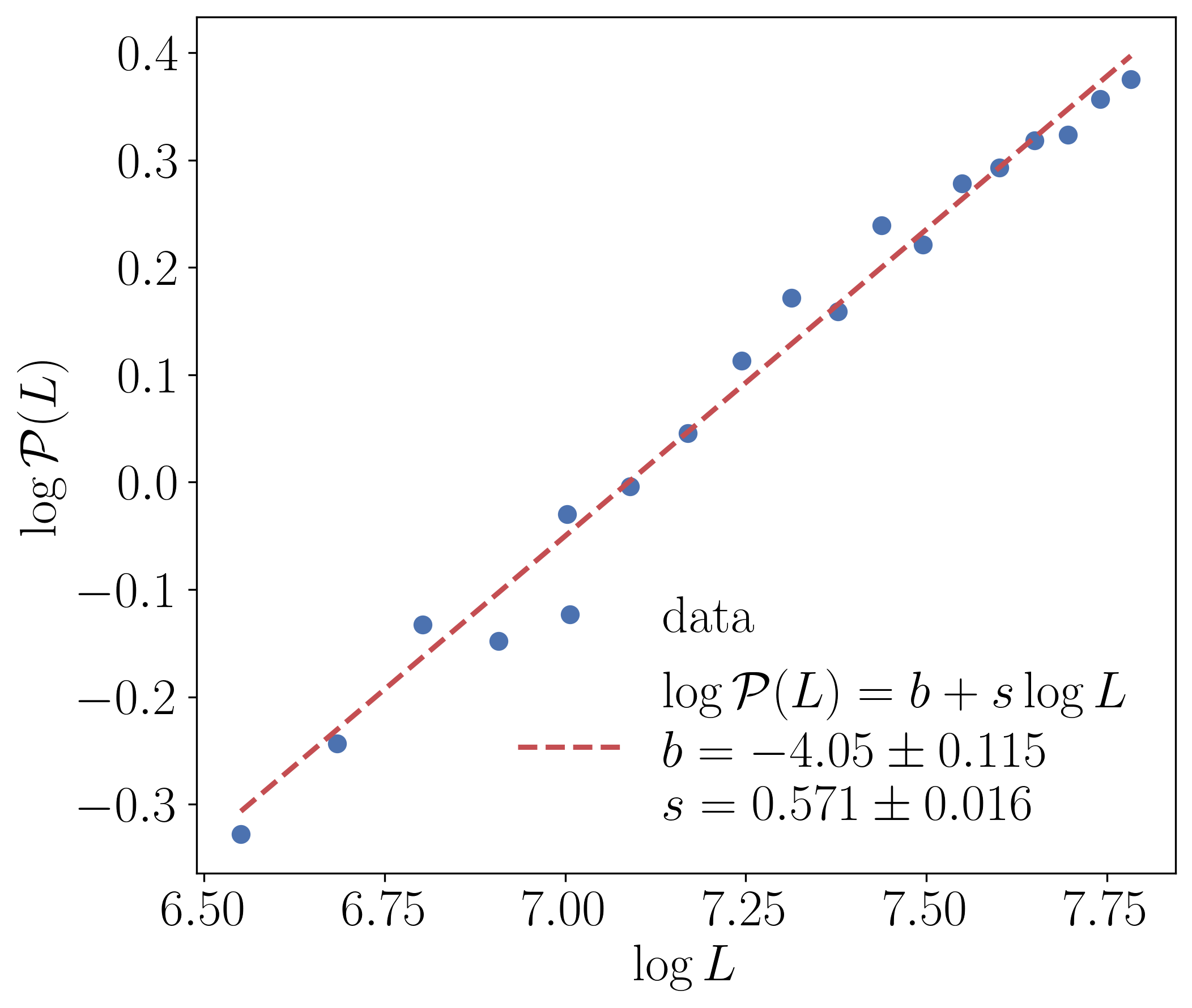}%
        }%
        \put(-170,143){(b)}%
        \put(-95,130){$\alpha=0.6$}%
    }\\
    \hfill
    \makebox[0.4\textwidth][c]{%
        \raisebox{0pt}{%
            \includegraphics[width=0.7\columnwidth]{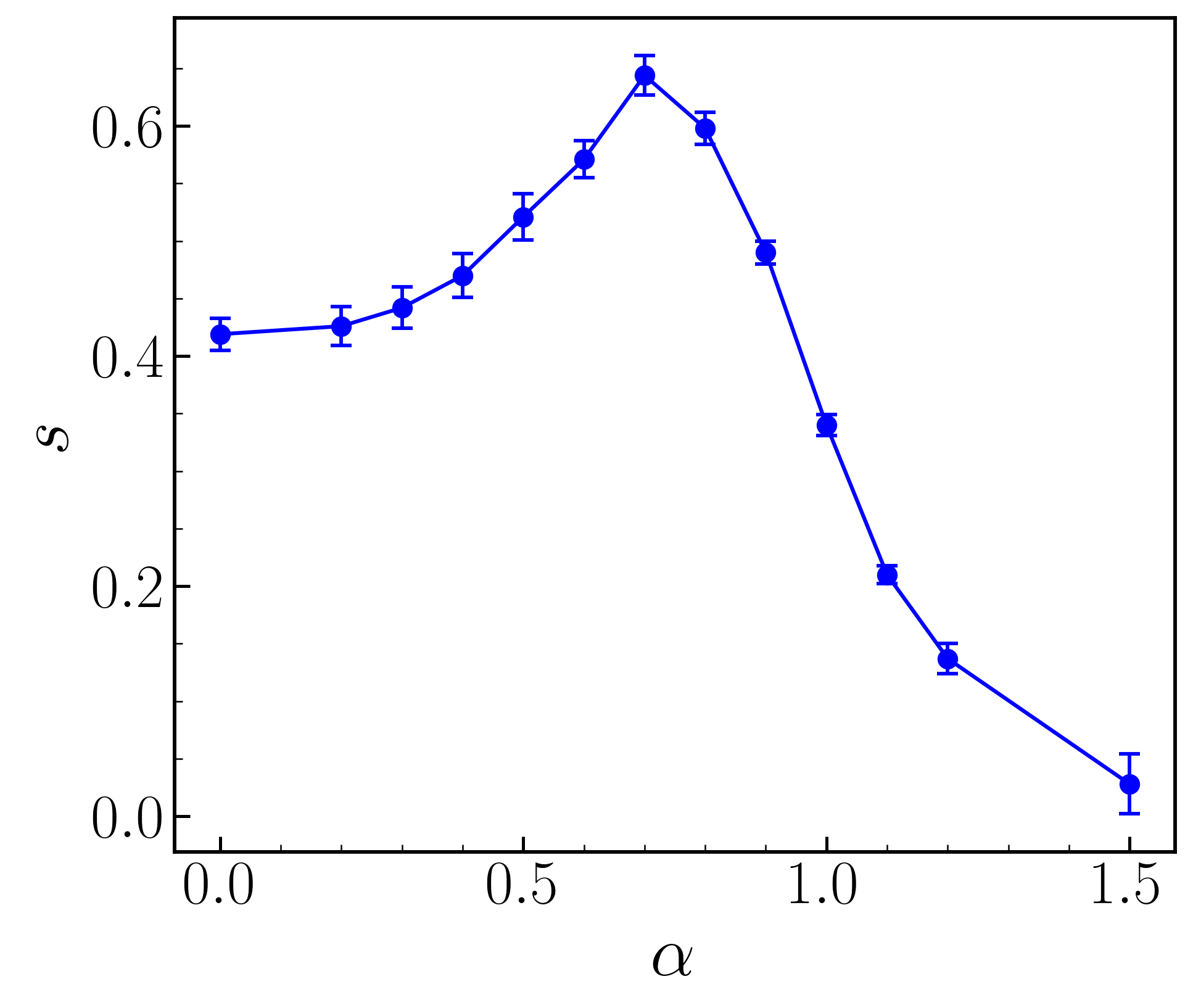}%
        }%
        \put(-165,143){(c)}%
    }
\end{tabular}
   \caption{
\textbf{Finite-size scaling of the Fourier-peak height.}
  Panels (a) and (b) show representative log--log fits for two selected
    parameter values. Symbols denote the numerical data, while dashed lines
    indicate the corresponding power-law fits. Error bars represent
    one-standard-deviation fit uncertainties. Panel (c) shows the correction exponent as the interaction-range parameter is varied.
Model parameters: \(\Omega=4.0\), \(\Delta=-8.9\), \(E=4.0\), and \(\chi=16\).
}
    \label{fig:FT_peak}
\end{figure}

For $\alpha=0$, the data shown in Fig.~\ref{fig:FT_peak}(a) display a clear
approximately linear behavior on logarithmic axes. The fit gives
\begin{equation}
s(0)=0.46\pm0.02.
\end{equation}
The positive value of $s(0)$ indicates that the Fourier-peak power increases
algebraically with system size over the range of sizes considered. We find a similar linear behavior for all the considered values of $\alpha$, and as an examples we show the case $\alpha=0.6$ in Fig.~\ref{fig:FT_peak}(b).

As $\alpha$ is increased, the extracted slope shows a nonmonotonic behavior, increasing up to $\alpha = 0.7$ and then decreasing and approaching
zero around $\alpha=1.5$ [see Fig.~\ref{fig:FT_peak}(c)]. A value
$s(\alpha)\simeq0$ indicates that the Fourier-peak power has only a weak
system-size dependence over the accessible range. The scaling exponent $s(\alpha)$ we get is positive for $\alpha\leq 1.2$, setting the boundary of the time-crystal phase at a value of $\alpha$ in the interval $[1,1.5]$ in full agreement with~\cite{Wang2025BoundaryTimeCrystals}. 
%
\section{Conclusion}\label{conc:sec}
We have developed a {semiclassical Langevin} approach to study finite-size effects
in dissipative time-crystalline spin systems. Starting from the quantum
Langevin equation associated with the Lindblad master equation, we derived
stochastic equations for local spin variables. This formulation retains the
deterministic mean-field dynamics while incorporating stochastic corrections
associated with the dissipative channels and is simpler than the Langevin DTWA approach~\cite{1wwv-k7hg,PhysRevA-105-013716}, where there is also a sampling over initial conditions, but the two approximations provide in our models identical results..

For the spin-\(1/2\) model with power-law Lindblad operators, we recover the
expected boundary-time-crystal behavior in the long-range regime. In the fully
collective limit, the stochastic dynamics approaches the mean-field oscillatory
solution as the system size increases. For finite power-law exponent \({\alpha}\),
the oscillation envelope decays exponentially at finite size, but the decay rate
decreases algebraically with system size. The corresponding scaling exponent
decreases as the Lindblad profile becomes shorter ranged and becomes compatible
with zero when the dissipative coupling is too short ranged. This provides a
direct finite-size characterization of the time-crystal lifetime. 

Remarkably, the threshold of $\alpha$ where the time crystal disappears lies above $\alpha = 1$. Therefore there is a range of $\alpha > 1$ where we predict the existence of a time crystal, in contrast with the relaxation to a fixed point predicted by the mean-field analysis in this range. Quantum fluctuations are relevant in the thermodynamic limit for $\alpha > 1$, then the mean field does not correctly describe the dynamics. Our approach takes into account at least part of the quantum correlations and provides a different result.


We also applied the same numerical approach to a spin-one model with local dissipation and
long-range Hamiltonian interactions. In this model the dissipation is strictly local
and the long-range coupling enters coherently through the Hamiltonian. In this model the thermodynamic-limit behavior is known to be described by a mean-field dynamics when $\alpha$ -- the exponent of the power-law interaction -- is smaller than 1. For that reason, in this range of $\alpha$, we can assess the persistence of the time-crystal oscillations in the thermodynamic limit by studying the system-size scaling of the
deviation time from the mean-field trajectory. As another probe for the presence of a time-crystal behavior, we also study the scaling with the system size of the dominant peak in the Fourier spectrum.

Both quantities grow algebraically with the system size in the fully connected
and long-range regimes, showing that the trajectory-averaged finite-size
dynamics remains close to a limit cycle for progressively longer
times. There is a range of $\alpha$ where both exponents are positive, marking the presence of a dissipative time crystals, and the one of the Fourier peak shows a nonmonotonic behavior in $\alpha$. As the coherent interaction becomes shorter ranged, both
exponents decrease and become close to zero for $\alpha > 1.2$, marking the disappearance of the time crystal.

The threshold $\alpha \simeq 1.2$ we find for the time crystal behavior lies in the interval $\alpha\in [1,1.5]$, in agreement with the findings of~\cite{Wang2025BoundaryTimeCrystals}. The results obtained with the scaling of the dominant Fourier peak are the most solid ones, because they do not rely on any assumption on the thermodynamic-limit behavior. This is not true for the deviation-time method
that works only in the range $\alpha\leq 1$, where the mean-field dynamics is known to describe thermodynamic-limit behavior~\cite{Wang2025BoundaryTimeCrystals}. So the threshold $\alpha=1.2$ we find with the latter method is probably a numerical artifact.
The nontrivial dependence on $\alpha$ of both scaling exponents provides us furthermore the  result that, although the thermodynamic-limit dynamics for $\alpha\leq 1$ is mean-field, the finite-size scaling towards this limit is not. 

Overall, our results show that semiclassical Langevin dynamics provides a useful
probe of dissipative time crystals beyond deterministic mean-field theory. It
reproduces the mean-field oscillatory behavior in the appropriate thermodynamic
limit and quantifies how persistence emerges with increasing system size. The
decay-rate exponent for the spin-\(1/2\) model and the deviation-time exponent
for the spin-one model provide complementary measures of the robustness of the
time-crystalline phase, especially in regimes where exact Lindblad simulations
are limited to small systems.

\begin{acknowledgements}
 R.K. gratefully acknowledges the resources on the LiCCA HPC cluster of the University of Augsburg, co-funded by the Deutsche Forschungsgemeinschaft (DFG, German Research Foundation)–Project-ID 499211671. A.\,R. acknowledges support from PNRR MUR Project PE0000023-NQSTI.
 \end{acknowledgements}



\appendix
\renewcommand{\theequation}{\thesection.\arabic{equation}}
\setcounter{equation}{0}

\section{Derivation of the spin-$1/2$ Langevin equations}
\label{app:spinhalf_langevin_derivation}

In this appendix we derive the semiclassical stochastic equations used in
Sec.~\ref{sec:spinhalf_model}. We start from the quantum Langevin equation,
Eq.~\eqref{eq:QLE}, with the power-law Lindblad operators
\begin{equation}
  {\hat{L}_j}=\sum_{l=1}^{L} f_{jl}({\alpha})\,\sigma_l^+ .
\end{equation}
Choosing the local operators \(\hat{O}=\sigma_i^x,\sigma_i^y,\sigma_i^z\), one obtains
\begin{align}
\dot{\sigma}_i^x
&=
-\frac{\gamma}{2}\sum_{j=1}^{L} f_{ji}^2({\alpha})\,\sigma_i^x
-\frac{\gamma}{4}
	\sum_{jl}
f_{ji}({\alpha})f_{jl}({\alpha})
\{\sigma_i^z,\sigma_l^x\}
\nonumber\\
&\quad
-\sqrt{\gamma}
\sum_{j=1}^{L} f_{ji}({\alpha})
\left[
\sigma_i^z\hat F_j(t)
+
\hat F_j^\dagger(t)\sigma_i^z
\right],
\nonumber\\[0.4em]
\dot{\sigma}_i^y
&=
-2J\sigma_i^z
-\frac{\gamma}{2}\sum_{j=1}^{L} f_{ji}^2({\alpha})\,\sigma_i^y
\nonumber\\
&\quad
-\frac{\gamma}{4}
\sum_{jl}
f_{ji}({\alpha})f_{jl}({\alpha})
\{\sigma_i^z,\sigma_l^y\}
\nonumber\\
&\quad
+\sqrt{\gamma}
\sum_{j=1}^{L} f_{ji}({\alpha})
\left[
i\sigma_i^z\hat F_j(t)
-
i\hat F_j^\dagger(t)\sigma_i^z
\right],
\nonumber\\[0.4em]
\dot{\sigma}_i^z
&=
2J\sigma_i^y
+
\gamma\sum_{j=1}^{L} f_{ji}^2({\alpha})(1-\sigma_i^z)
\nonumber\\
&\quad
+\frac{\gamma}{2}
\sum_{jl}
f_{ji}({\alpha})f_{jl}({\alpha})
\left(
\sigma_i^x\sigma_l^x
+
\sigma_i^y\sigma_l^y
\right)
\nonumber\\
&\quad
+2\sqrt{\gamma}
\sum_{j=1}^{L} f_{ji}({\alpha})
\left[
\sigma_i^-\hat F_j(t)
+
\hat F_j^\dagger(t)\sigma_i^+
\right].
\label{eq:spinhalf_operator_langevin}
\end{align}
The noise operators are decomposed as
\begin{equation}
  \hat F_j(t)
  =
  \frac{1}{2}
  \left[
  \xi_j^{(1)}(t)
  +
  i\xi_j^{(2)}(t)
  \right],
\end{equation}
where the real Gaussian noises satisfy
\begin{equation}
  \left\langle
  \xi_j^{(\alpha)}(t)
  \xi_l^{(\beta)}(t')
  \right\rangle
  =
  \delta_{jl}
  \delta_{\alpha\beta}
  \delta(t-t') .
\end{equation}
Substitution gives
\begin{align}
\dot{\sigma}_i^x
&=
-\frac{\gamma}{2}
\sum_{j=1}^{L} f_{ji}^2({\alpha})\,\sigma_i^x
-\frac{\gamma}{4}
\sum_{jl}
f_{ji}({\alpha})f_{jl}({\alpha})
\{\sigma_i^z,\sigma_l^x\}
\nonumber\\
&\quad
-\sqrt{\gamma}
\sum_{j=1}^{L}
f_{ji}({\alpha})\xi_j^{(1)}(t)\sigma_i^z,
\nonumber\\[0.4em]
\dot{\sigma}_i^y
&=
-2J\sigma_i^z
-\frac{\gamma}{2}
\sum_{j=1}^{L} f_{ji}^2({\alpha})\,\sigma_i^y
-\frac{\gamma}{4}
\sum_{jl}
f_{ji}({\alpha})f_{jl}({\alpha})
\{\sigma_i^z,\sigma_l^y\}
\nonumber\\
&\quad
-\sqrt{\gamma}
\sum_{j=1}^{L}
f_{ji}({\alpha})\xi_j^{(2)}(t)\sigma_i^z,
\nonumber\\[0.4em]
\dot{\sigma}_i^z
&=
2J\sigma_i^y
+
\gamma
\sum_{j=1}^{L}
f_{ji}^2({\alpha})(1-\sigma_i^z)
\nonumber\\
&\quad
+\frac{\gamma}{2}
\sum_{jl}
f_{ji}({\alpha})f_{jl}({\alpha})
\left(
\sigma_i^x\sigma_l^x
+
\sigma_i^y\sigma_l^y
\right)
\nonumber\\
&\quad
+\sqrt{\gamma}
\sum_{j=1}^{L} f_{ji}({\alpha})
\left[
\xi_j^{(1)}(t)\sigma_i^x
+
\xi_j^{(2)}(t)\sigma_i^y
\right].
\label{eq:spinhalf_operator_real_noise}
\end{align}
We then introduce the semiclassical variables
\begin{equation}
  s_i^\alpha(t)=\langle\sigma_i^\alpha(t)\rangle,
  \qquad
  \alpha=x,y,z,
\end{equation}
and factorize operator products as
\begin{equation}
  \langle \sigma_i^\alpha\sigma_l^\beta\rangle
  \simeq
  s_i^\alpha s_l^\beta .
\end{equation}
For operators on different sites this also gives
\begin{equation}
  \langle\{\sigma_i^z,\sigma_l^\alpha\}\rangle
  \simeq
  2s_i^z s_l^\alpha,
  \qquad
  l\neq i .
\end{equation}
Finally, the white noises are written in terms of Wiener increments as
\begin{equation}
  \xi_j^{(\alpha)}(t)\ud t
  =
  \ud W_j^{(\alpha)} .
\end{equation}
To connect with the compact notation used in the main text, we define the
effective dissipative coupling matrix
\begin{equation}
G_{il}({\alpha})
=
\sum_{j=1}^{L}
f_{ji}({\alpha})f_{jl}({\alpha}),
\end{equation}
and the corresponding noise increments
\begin{equation}
d\mathcal W_i^{(\alpha)}
=
\sum_{j=1}^{L}
f_{ji}({\alpha})dW_j^{(\alpha)} .
\end{equation}
These increments satisfy
\begin{equation}
\left\langle
d\mathcal W_i^{(\alpha)}
d\mathcal W_l^{(\beta)}
\right\rangle
=
\delta_{\alpha\beta}
G_{il}({\alpha})\,dt .
\end{equation}
We also introduce the nonlocal transverse fields
\begin{equation}
h_i^\mu(t)
=
\sum_{l\neq i}
G_{il}({\alpha})s_l^\mu(t),
\qquad
\mu=x,y .
\end{equation}
With these definitions, the expanded semiclassical equations reduce to the
compact Itô equations reported in
Eq.~\eqref{eq:spinhalf_semiclassical_sde_compact}.

\begin{figure}
    \centering
    \includegraphics[width=0.7\columnwidth]{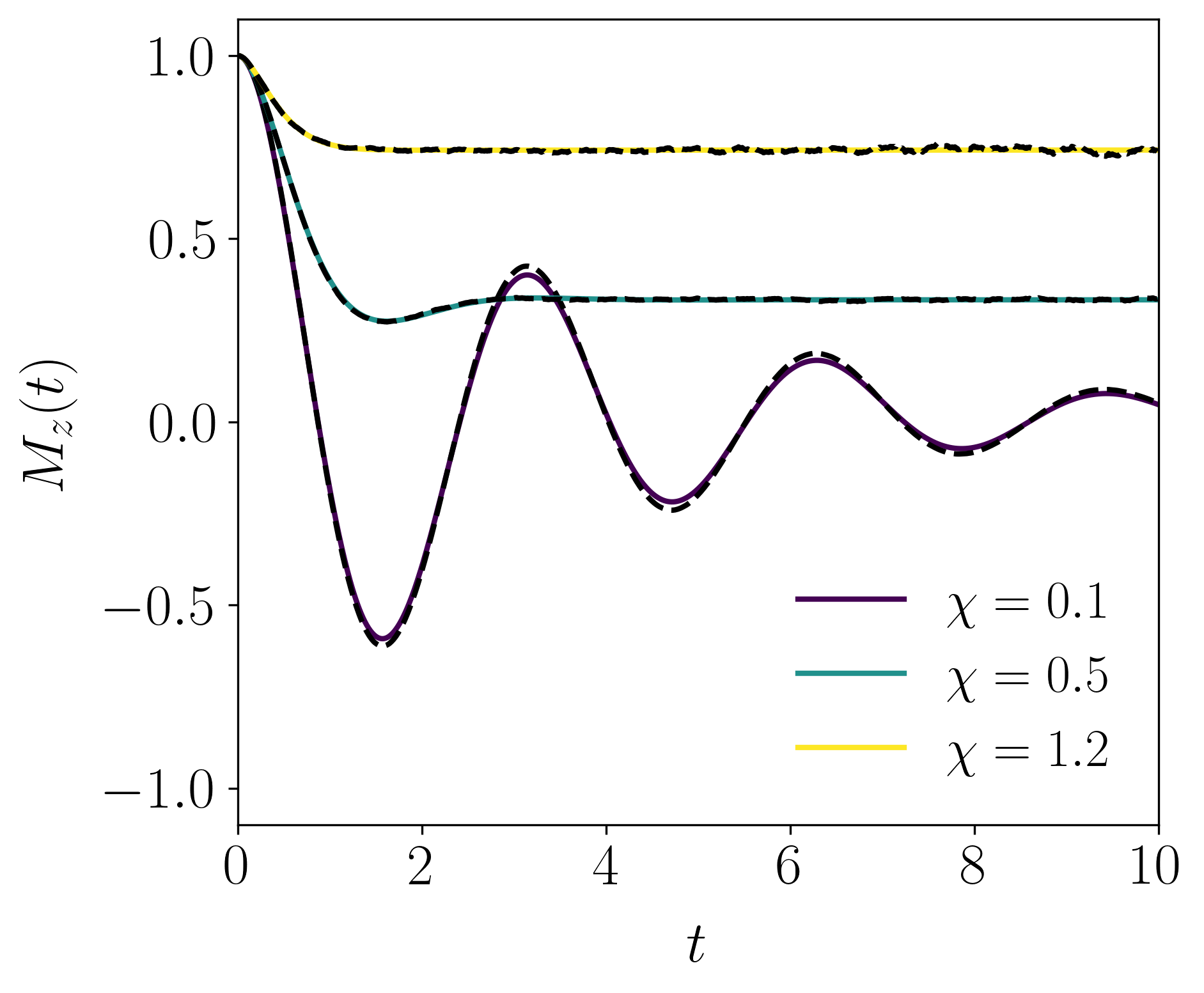}\put(-150,150){(a)}\\
    \includegraphics[width=0.7\columnwidth]{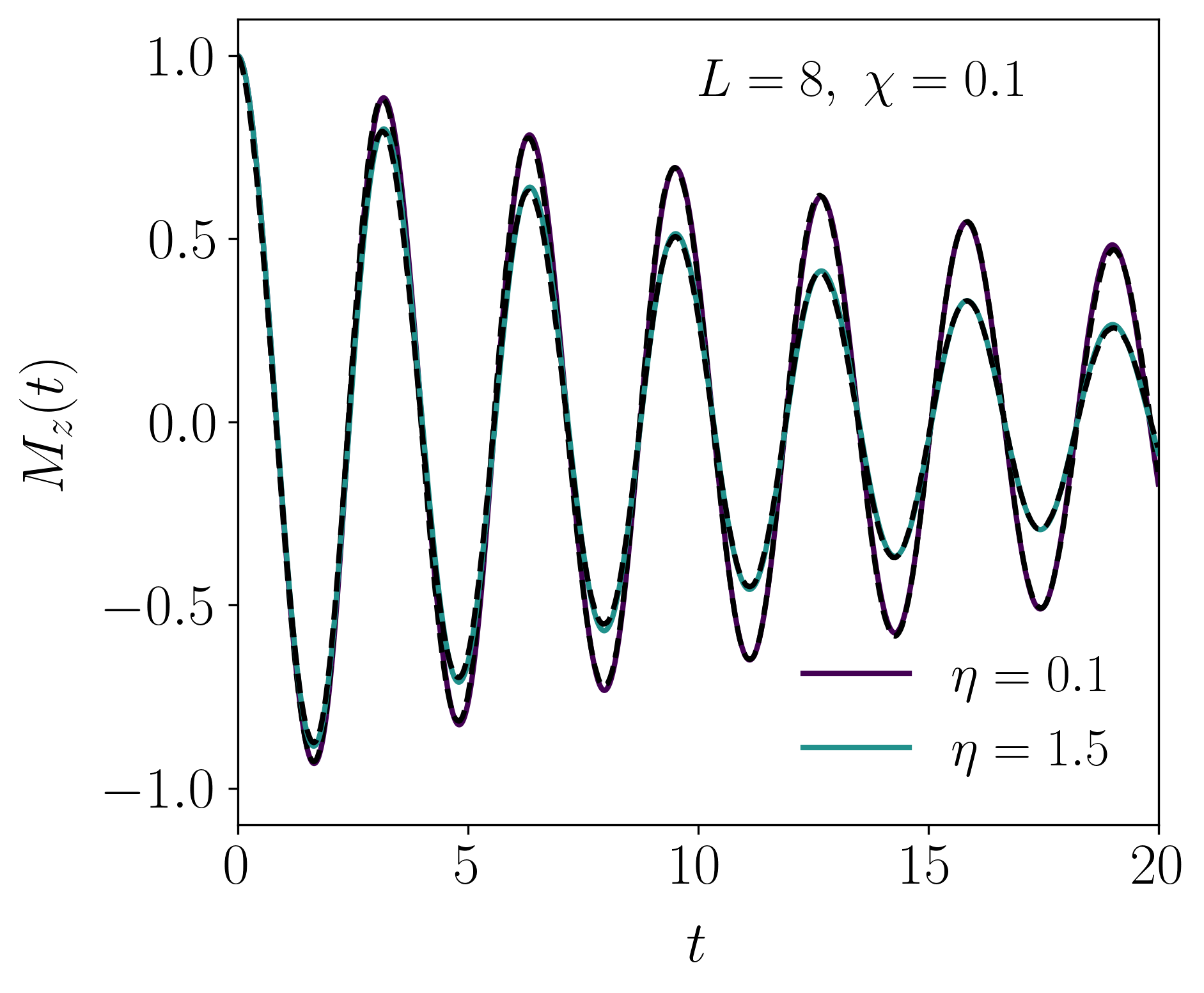}\put(-150,150){(b)}
    \caption{
    \textbf{Semiclassical versus exact Lindblad dynamics.}
    (a) Dynamics in the local limit \({\alpha}\to\infty\), where the model reduces
    to a single spin coupled to a Markovian bath. We compare
    \(\langle\sigma^z(t)\rangle\) obtained from the trajectory-averaged
    semiclassical Langevin dynamics (dashed lines) with the exact solution of
    the Lindblad master equation (solid lines) for \(\chi=0.1\), \(0.5\), and
    \(1.2\).
    (b) Dynamics for finite power-law Lindblad operators in a system of size
    \(L=8\) at \(\chi=0.1\). The trajectory-averaged Langevin dynamics
    (dashed lines) is compared with the exact many-body Lindblad evolution
    (solid lines) for \({\alpha}=0.1\) and \({\alpha}=1.5\).
    }
    \label{fig:sigma_z_comparison_rungekutta}
\end{figure}

\section{Spin-$1/2$ model: comparison with exact diagonalization results}
\label{supp:s1}

In this section we benchmark the semiclassical dynamics generated by the
Langevin equations against the exact quantum evolution governed by the Lindblad
master equation. This comparison is useful for validating the stochastic
equations in regimes where exact solutions are still accessible.

We first consider the limit \({\alpha}\to\infty\), in which the spatial profile of
the Lindblad operators becomes local,
\begin{equation}
f_{jl}({\alpha})\to\delta_{jl}.
\end{equation}
In this regime the sites are decoupled, and the many-body dynamics reduces to
the evolution of a single spin coupled to a Markovian bath. The corresponding
stochastic differential equations for the spin components
\(s^\beta\), with \(\beta=x,y,z\), are
\begin{align}
  \ud s^x
  &=
  -\frac{\gamma}{2}s^x\,\ud t
  -
  \sqrt{\gamma}\,s^z\,\ud W^{(1)},
  \nonumber\\
  \ud s^y
  &=
  -2J\,s^z\,\ud t
  -
  \frac{\gamma}{2}s^y\,\ud t
  -
  \sqrt{\gamma}\,s^z\,\ud W^{(2)},
  \nonumber\\
  \ud s^z
  &=
  2J\,s^y\,\ud t
  +
  \gamma(1-s^z)\,\ud t
  +
  \sqrt{\gamma}
  \left(
  s^x\,\ud W^{(1)}
  +
  s^y\,\ud W^{(2)}
  \right).
\end{align}
Figure~\ref{fig:sigma_z_comparison_rungekutta}(a) shows
\(\langle\sigma^z(t)\rangle\) for \(\chi=0.1\), \(0.5\), and \(1.2\). The dashed
curves are obtained from the trajectory-averaged semiclassical Langevin
dynamics, while the solid curves are obtained from the exact Lindblad evolution
of the single-spin density matrix. The agreement is essentially exact for all
three values of \(\chi\), including both the oscillatory regime and the
overdamped relaxation regime. This confirms that, in the strictly local limit,
the Langevin formulation reproduces the exact open-system dynamics.

We next consider finite-range power-law Lindblad operators, with
\(f_{jl}({\alpha})\) defined in Eq.~\eqref{eq:power-law-lindblad}. In this case
different spins are coupled through the dissipative channels, and the
corresponding semiclassical dynamics is governed by the many-body Langevin
equations in Eq.~\eqref{eq:spinhalf_semiclassical_sde_compact}. We solve these
coupled stochastic differential equations and compare the trajectory-averaged
magnetization with the exact many-body Lindblad evolution for small system sizes.

Figure~\ref{fig:sigma_z_comparison_rungekutta}(b) shows this comparison for
\(L=8\), \(\chi=0.1\), and two values of the power-law exponent,
\({\alpha}=0.1\) and \({\alpha}=1.5\). Again, dashed curves denote the semiclassical
Langevin results and solid curves denote the exact Lindblad dynamics. The
agreement is very good over the simulated time window. The comparison also shows
that changing \({\alpha}\) modifies the damping of the oscillations: the
shorter-ranged case \({\alpha}=1.5\) exhibits stronger damping than the more
long-ranged case \({\alpha}=0.1\). This benchmark supports the use of the
semiclassical Langevin equations to study finite-size decay and long-range
time-crystalline behavior beyond the system sizes accessible to exact
diagonalization.


\begin{table*}[t]
\centering
\scriptsize
\setlength{\tabcolsep}{2.2pt}
\renewcommand{\arraystretch}{1.12}

\begin{adjustbox}{max width=\textwidth}
\begin{tabular}{>{\columncolor{FirstColBlue}}c cccccccc}
\rowcolor{HeaderBlue}
\multicolumn{9}{c}{\textbf{(a) Commutation relations}}\\
\toprule
\rowcolor{HeaderBlue}
$[\,,\,]$
& $\Lambda_1$ & $\Lambda_2$ & $\Lambda_3$ & $\Lambda_4$
& $\Lambda_5$ & $\Lambda_6$ & $\Lambda_7$ & $\Lambda_8$ \\
\midrule
$\Lambda_1$
& \emptycell
& $i\Lambda_6/\sqrt{2}$
& $i\Lambda_5/\sqrt{2}$
& $i\sqrt{2}\Lambda_7$
& $-i\Lambda_3/\sqrt{2}$
& $-i\Lambda_2/\sqrt{2}$
& $-i\sqrt{2}\Lambda_4$
& $0$ \\

$\Lambda_2$
& \emptycell
& \emptycell
& $i\Lambda_4/\sqrt{2}$
& $-i\Lambda_3/\sqrt{2}$
& $i(\Lambda_7/\sqrt{2}+\sqrt{3/2}\Lambda_8)$
& $i\Lambda_1/\sqrt{2}$
& $-i\Lambda_5/\sqrt{2}$
& $-i\sqrt{3/2}\Lambda_5$ \\

$\Lambda_3$
& \emptycell
& \emptycell
& \emptycell
& $i\Lambda_2/\sqrt{2}$
& $i\Lambda_1/\sqrt{2}$
& $\frac{i}{2}(\sqrt{6}\Lambda_8-\sqrt{2}\Lambda_7)$
& $i\Lambda_6/\sqrt{2}$
& $-i\sqrt{3/2}\Lambda_6$ \\

$\Lambda_4$
& \emptycell
& \emptycell
& \emptycell
& \emptycell
& $i\Lambda_6/\sqrt{2}$
& $-i\Lambda_5/\sqrt{2}$
& $i\sqrt{2}\Lambda_1$
& $0$ \\

$\Lambda_5$
& \emptycell
& \emptycell
& \emptycell
& \emptycell
& \emptycell
& $i\Lambda_4/\sqrt{2}$
& $i\Lambda_2/\sqrt{2}$
& $i\sqrt{3/2}\Lambda_2$ \\

$\Lambda_6$
& \emptycell
& \emptycell
& \emptycell
& \emptycell
& \emptycell
& \emptycell
& $-i\Lambda_3/\sqrt{2}$
& $i\sqrt{3/2}\Lambda_3$ \\

$\Lambda_7$
& \emptycell
& \emptycell
& \emptycell
& \emptycell
& \emptycell
& \emptycell
& \emptycell
& $0$ \\
\bottomrule
\end{tabular}
\end{adjustbox}

\vspace{0.7em}

\begin{adjustbox}{max width=\textwidth}
\begin{tabular}{>{\columncolor{FirstColBlue}}c cccccccc}
\rowcolor{HeaderBlue}
\multicolumn{9}{c}{\textbf{(b) Anticommutation relations}}\\
\toprule
\rowcolor{HeaderBlue}
$\{\,,\,\}$
& $\Lambda_1$ & $\Lambda_2$ & $\Lambda_3$ & $\Lambda_4$
& $\Lambda_5$ & $\Lambda_6$ & $\Lambda_7$ & $\Lambda_8$ \\
\midrule
$\Lambda_1$
& \emptycell
& $\Lambda_3/\sqrt{2}$
& $\Lambda_2/\sqrt{2}$
& $0$
& $\Lambda_6/\sqrt{2}$
& $\Lambda_5/\sqrt{2}$
& $0$
& $\sqrt{2/3}\Lambda_1$ \\

$\Lambda_2$
& \emptycell
& \emptycell
& $\Lambda_1/\sqrt{2}$
& $-\Lambda_6/\sqrt{2}$
& $0$
& $-\Lambda_4/\sqrt{2}$
& $\Lambda_2/\sqrt{2}$
& $-\Lambda_2/\sqrt{6}$ \\

$\Lambda_3$
& \emptycell
& \emptycell
& \emptycell
& $\Lambda_5/\sqrt{2}$
& $\Lambda_4/\sqrt{2}$
& $0$
& $-\Lambda_3/\sqrt{2}$
& $-\Lambda_3/\sqrt{6}$ \\

$\Lambda_4$
& \emptycell
& \emptycell
& \emptycell
& \emptycell
& $\Lambda_3/\sqrt{2}$
& $-\Lambda_2/\sqrt{2}$
& $0$
& $\sqrt{2/3}\Lambda_4$ \\

$\Lambda_5$
& \emptycell
& \emptycell
& \emptycell
& \emptycell
& \emptycell
& $\Lambda_1/\sqrt{2}$
& $\Lambda_5/\sqrt{2}$
& $-\Lambda_5/\sqrt{6}$ \\

$\Lambda_6$
& \emptycell
& \emptycell
& \emptycell
& \emptycell
& \emptycell
& \emptycell
& $-\Lambda_6/\sqrt{2}$
& $-\Lambda_6/\sqrt{6}$ \\

$\Lambda_7$
& \emptycell
& \emptycell
& \emptycell
& \emptycell
& \emptycell
& \emptycell
& \emptycell
& $\sqrt{2/3}\Lambda_7$ \\
\bottomrule
\end{tabular}
\end{adjustbox}

\caption{
\textbf{Algebra of the normalized Gell-Mann basis.}
(a) Commutation relations and (b) anticommutation relations. Only
upper-triangular entries are shown. Lower-triangular commutators follow from
antisymmetry, while lower-triangular anticommutators follow from symmetry.
Identity contributions in diagonal anticommutators are omitted.
Here \(\Lambda_\mu\equiv\hat{\Lambda}_\mu\).
}
\label{tab:gellmann-algebra}
\end{table*}
\section{Algebra of the Gell-Mann basis}
\label{app:gellmann-algebra}

In this appendix we collect the algebraic relations used to derive the
Heisenberg-Langevin equations for the spin-one model. The local three-level
Hilbert space is represented in terms of the eight Hermitian Gell-Mann matrices
\(\hat{\Lambda}_\mu\), normalized as
\begin{equation}
\mathrm{Tr}
\left(
\hat{\Lambda}_\mu\hat{\Lambda}_\nu
\right)
=
\delta_{\mu\nu}.
\end{equation}
Products of Gell-Mann matrices can be decomposed into commutators and
anticommutators,
\begin{equation}
\hat{\Lambda}_\mu\hat{\Lambda}_\nu
=
\frac{1}{2}
\left[
\hat{\Lambda}_\mu,\hat{\Lambda}_\nu
\right]
+
\frac{1}{2}
\left\{
\hat{\Lambda}_\mu,\hat{\Lambda}_\nu
\right\}.
\end{equation}
The commutators determine the coherent Hamiltonian contribution to the equations
of motion, while the anticommutators enter the dissipative terms generated by
the Lindblad operators. The nonzero independent entries are listed in
Tables~\ref{tab:gellmann-algebra}. The remaining entries follow from
antisymmetry of the commutator,
\begin{equation}
[\hat{\Lambda}_\mu,\hat{\Lambda}_\nu]
=
-
[\hat{\Lambda}_\nu,\hat{\Lambda}_\mu],
\end{equation}
and symmetry of the anticommutator,
\begin{equation}
\{\hat{\Lambda}_\mu,\hat{\Lambda}_\nu\}
=
\{\hat{\Lambda}_\nu,\hat{\Lambda}_\mu\}.
\end{equation}
For compactness, the identity contributions in anticommutators with identical
indices are not displayed in Table~\ref{tab:gellmann-algebra}, since the table is used only
for the operator products appearing in the Langevin derivation.

\section{Derivation of the spin-one Langevin equations}
\label{app:spinone_langevin_derivation}

In this appendix we give the operator-level derivation of the semiclassical
spin-one Langevin equations used in Sec.~\ref{subsec:spinone_semiclassical}.
Starting from the quantum Langevin equation, Eq.~\eqref{eq:QLE}, we apply it to
the local Gell-Mann operators \(\hat{\Lambda}_{\mu j}\). The  jump
operators are defined in Eq.~\eqref{lops1:eqn} and
%
%
%
%
their representation in the Gell-Mann basis is given in
Eq.~\eqref{eq:spinone_jump_gellmann}.

Using the commutation and anticommutation relations of the Gell-Mann matrices,
listed in Tables~\ref{tab:gellmann-algebra}, the operator
Heisenberg-Langevin equations become
\begin{widetext}
\begin{align}
\label{eq:spinone_operator_langevin}
\dot{\hat{\Lambda}}_{1j}
&=
\frac{\Omega}{2}\hat{\Lambda}_{5j}
+
(\Delta+E)\hat{\Lambda}_{4j}
+
\sum_{i<j}V_{ij}\hat{n}_i\hat{\Lambda}_{4j}
+
\sum_{i>j}V_{ij}\hat{\Lambda}_{4j}\hat{n}_i
-
\frac{\gamma}{2}\hat{\Lambda}_{1j}
\nonumber\\
&\quad
+
\sqrt{\gamma}
\left[
\hat{\Lambda}_{7j}\hat{F}_{+j}(t)
+
\frac12
\left(
\hat{\Lambda}_{2j}
-
i\hat{\Lambda}_{5j}
\right)
\hat{F}_{-j}(t)
+
\mathrm{H.c.}
\right],
\nonumber\\
\dot{\hat{\Lambda}}_{2j}
&=
\frac{\Omega}{2}
\left(
\hat{\Lambda}_{4j}
-
\hat{\Lambda}_{6j}
\right)
+
2E\hat{\Lambda}_{5j}
-
\gamma\hat{\Lambda}_{2j}
\nonumber\\
&\quad
+
\frac{\sqrt{\gamma}}{2}
\left[
\left(
\hat{\Lambda}_{3j}
-
i\hat{\Lambda}_{6j}
\right)
\hat{F}_{+j}(t)
+
\left(
\hat{\Lambda}_{1j}
+
i\hat{\Lambda}_{4j}
\right)
\hat{F}_{-j}(t)
+
\mathrm{H.c.}
\right],
\nonumber\\
\dot{\hat{\Lambda}}_{3j}
&=
-\frac{\Omega}{2}\hat{\Lambda}_{5j}
-
(\Delta-E)\hat{\Lambda}_{6j}
-
\sum_{i<j}V_{ij}\hat{n}_i\hat{\Lambda}_{6j}
-
\sum_{i>j}V_{ji}\hat{\Lambda}_{6j}\hat{n}_i
-
\frac{\gamma}{2}\hat{\Lambda}_{3j}
\nonumber\\
&\quad
+
\frac{\sqrt{\gamma}}{2}
\left[
\left(
\hat{\Lambda}_{2j}
+
i\hat{\Lambda}_{5j}
\right)
\hat{F}_{+j}(t)
-
\left(
\sqrt{3}\hat{\Lambda}_{8j}
-
\hat{\Lambda}_{7j}
\right)
\hat{F}_{-j}(t)
+
\mathrm{H.c.}
\right],
\nonumber\\
\dot{\hat{\Lambda}}_{4j}
&=
-\Omega
\left(
\hat{\Lambda}_{7j}
+
\frac12\hat{\Lambda}_{2j}
\right)
-
(\Delta+E)\hat{\Lambda}_{1j}
-
\sum_{i<j}V_{ij}\hat{n}_i\hat{\Lambda}_{1j}
-
\sum_{i>j}V_{ji}\hat{\Lambda}_{1j}\hat{n}_i
-
\frac{\gamma}{2}\hat{\Lambda}_{4j}
\nonumber\\
&\quad
+
\sqrt{\gamma}
\left[
i\hat{\Lambda}_{7j}\hat{F}_{+j}(t)
+
\frac12
\left(
\hat{\Lambda}_{5j}
+
i\hat{\Lambda}_{2j}
\right)
\hat{F}_{-j}(t)
+
\mathrm{H.c.}
\right],
\nonumber\\
\dot{\hat{\Lambda}}_{5j}
&=
-\frac{\Omega}{2}
\left(
\hat{\Lambda}_{1j}
-
\hat{\Lambda}_{3j}
\right)
-
2E\hat{\Lambda}_{2j}
-
\gamma\hat{\Lambda}_{5j}
\nonumber\\
&\quad
-
\frac{\sqrt{\gamma}}{2}
\left[
\left(
\hat{\Lambda}_{6j}
+
i\hat{\Lambda}_{3j}
\right)
\hat{F}_{+j}(t)
+
\left(
\hat{\Lambda}_{4j}
-
i\hat{\Lambda}_{1j}
\right)
\hat{F}_{-j}(t)
+
\mathrm{H.c.}
\right],
\nonumber\\
\dot{\hat{\Lambda}}_{6j}
&=
\Omega
\left(
\frac12\hat{\Lambda}_{2j}
-
\frac{\sqrt{3}}{2}\hat{\Lambda}_{8j}
+
\frac12\hat{\Lambda}_{7j}
\right)
+
(\Delta-E)\hat{\Lambda}_{3j}
+
\sum_{i<j}V_{ij}\hat{n}_i\hat{\Lambda}_{3j}
+
\sum_{i>j}V_{ij}\hat{\Lambda}_{3j}\hat{n}_i
-
\frac{\gamma}{2}\hat{\Lambda}_{6j}
\nonumber\\
&\quad
+
\frac{\sqrt{\gamma}}{2}
\left[
\left(
\hat{\Lambda}_{5j}
-
i\hat{\Lambda}_{2j}
\right)
\hat{F}_{+j}(t)
+
i
\left(
\sqrt{3}\hat{\Lambda}_{8j}
-
\hat{\Lambda}_{7j}
\right)
\hat{F}_{-j}(t)
+
\mathrm{H.c.}
\right],
\nonumber\\
\dot{\hat{\Lambda}}_{7j}
&=
\Omega\hat{\Lambda}_{4j}
-
\frac{\Omega}{2}\hat{\Lambda}_{6j}
-
\frac{\gamma}{\sqrt{2}}\hat{\boldsymbol{1}}_j
-
\gamma\hat{\Lambda}_{7j}
\nonumber\\
&\quad
-
\sqrt{\gamma}
\left[
\left(
\hat{\Lambda}_{1j}
+
i\hat{\Lambda}_{4j}
\right)
\hat{F}_{+j}(t)
+
\frac12
\left(
\hat{\Lambda}_{3j}
-
i\hat{\Lambda}_{6j}
\right)
\hat{F}_{-j}(t)
+
\mathrm{H.c.}
\right],
\nonumber\\
\dot{\hat{\Lambda}}_{8j}
&=
\frac{\sqrt{3}}{2}\Omega\hat{\Lambda}_{6j}
+
\frac{\gamma}{\sqrt{6}}\hat{\boldsymbol{1}}_j
-
\gamma\hat{\Lambda}_{8j}
+
\sqrt{\frac{3\gamma}{2}}
\left[
\left(
\hat{\Lambda}_{3j}
-
i\hat{\Lambda}_{6j}
\right)
\hat{F}_{-j}(t)
+
\mathrm{H.c.}
\right].
\end{align}
\end{widetext}
The noise operators are decomposed as
\begin{equation}
  \hat{F}_{\sigma j}(t)
  =
  \frac12
  \left[
  \xi_{\sigma j}^{(1)}(t)
  +
  i\xi_{\sigma j}^{(2)}(t)
  \right],
\end{equation}
where the real noises satisfy
\begin{equation}
  \left\langle
  \xi_{\sigma j}^{(a)}(t)
  \xi_{\tau l}^{(b)}(t')
  \right\rangle
  =
  \delta^{ab}
  \delta_{\sigma\tau}
  \delta_{jl}
  \delta(t-t') .
\end{equation}
We then replace the operators \(\hat{\Lambda}_{\mu j}\) by the semiclassical
variables
\begin{equation}
  \lambda_{\mu j}(t)
  =
  \left\langle
  \hat{\Lambda}_{\mu j}(t)
  \right\rangle ,
\end{equation}
and factorize products on different sites. The interaction term is expressed
through
\begin{equation}
  n_i
  =
  \frac{2}{3}
  -
  \frac{\sqrt{6}}{6}\lambda_{8i}
  +
  \frac{\sqrt{2}}{2}\lambda_{7i},
\end{equation}
which follows from
\begin{equation}
  \hat{n}_i
  =
  (\hat{s}_i^z)^2
  =
  \frac{2}{3}\boldsymbol{1}
  -
  \frac{\sqrt{6}}{6}\hat{\Lambda}_{8i}
  +
  \frac{\sqrt{2}}{2}\hat{\Lambda}_{7i}.
\end{equation}
Using the symmetry \(V_{ij}=V_{ji}\), the interaction contribution can be
written compactly by defining the interaction-shifted detuning
\begin{equation}
  \Delta_j(t)
  =
  \Delta
  +
  \sum_{i\neq j}V_{ij}n_i(t).
\end{equation}
We also introduce the shorthand
\begin{equation}
  q_j(t)
  =
  \sqrt{3}\lambda_{8j}(t)
  -
  \lambda_{7j}(t).
\end{equation}
With these definitions, the semiclassical Langevin equations take the compact
form used in the main text,
\begin{equation}
\label{eq:spinone_sde_appendix}
  \dot{\lambda}_{\mu j}
  =
  A_{\mu j}(\boldsymbol{\lambda})
  +
  \sqrt{\gamma}\,
  D_{\mu j}(\boldsymbol{\lambda},\boldsymbol{\xi}),
  \qquad
  \mu=1,\ldots,8 .
\end{equation}
The deterministic drift terms are
\begin{figure}
    \centering
    \includegraphics[width=80mm]{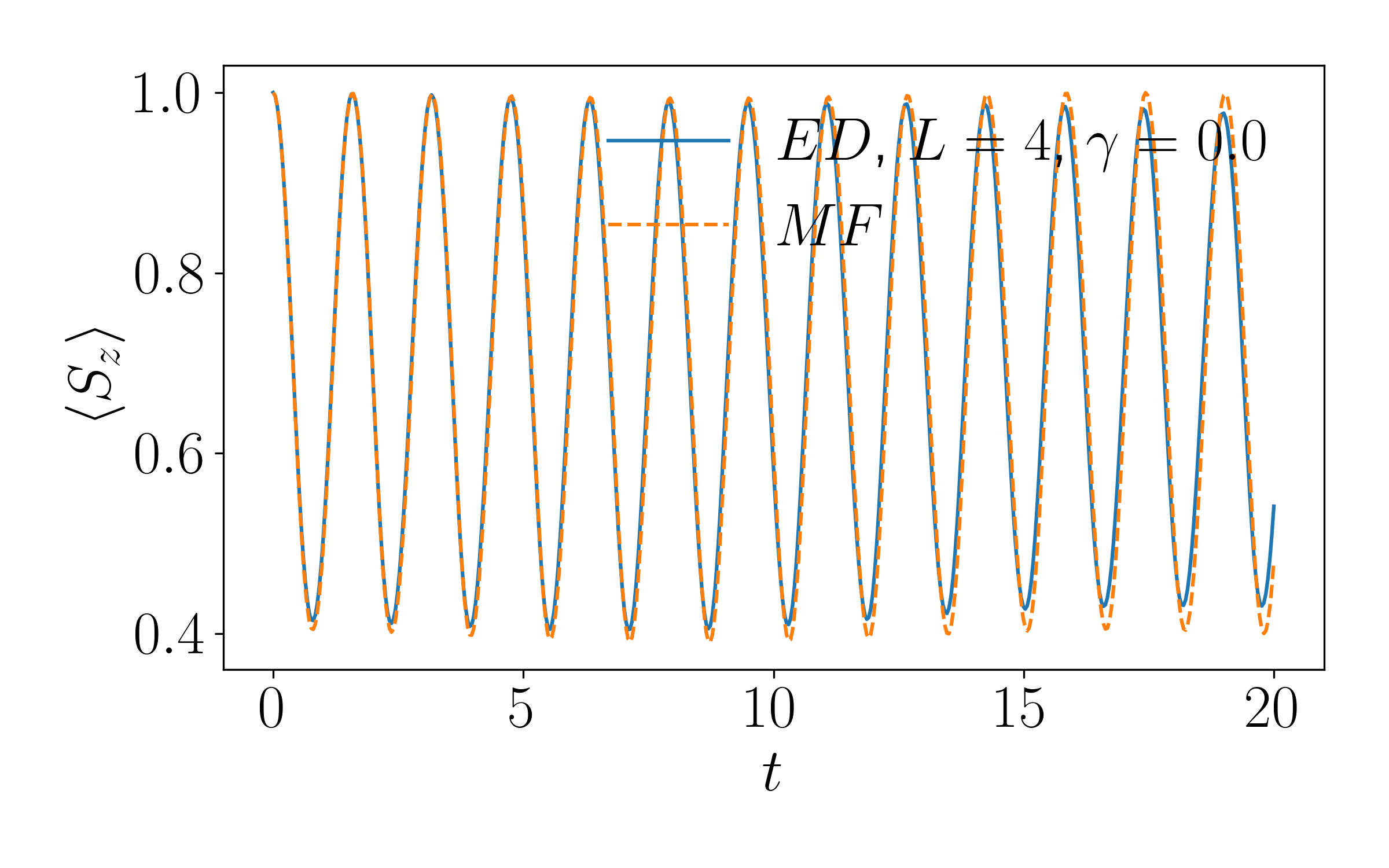}\put(-120,90){}\\
    \includegraphics[width=80mm]{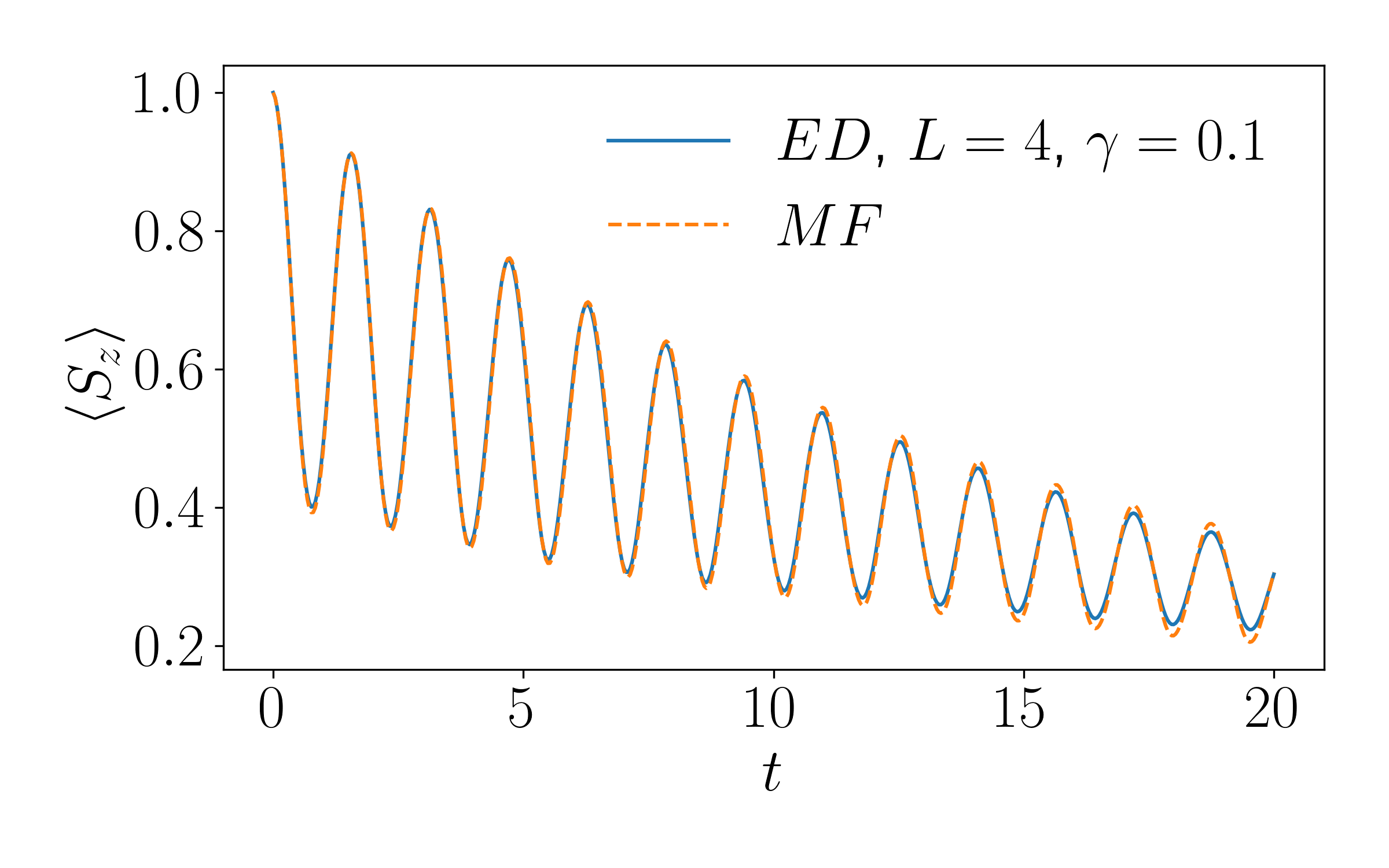}\put(-120,90){}\\
    \includegraphics[width=80mm]{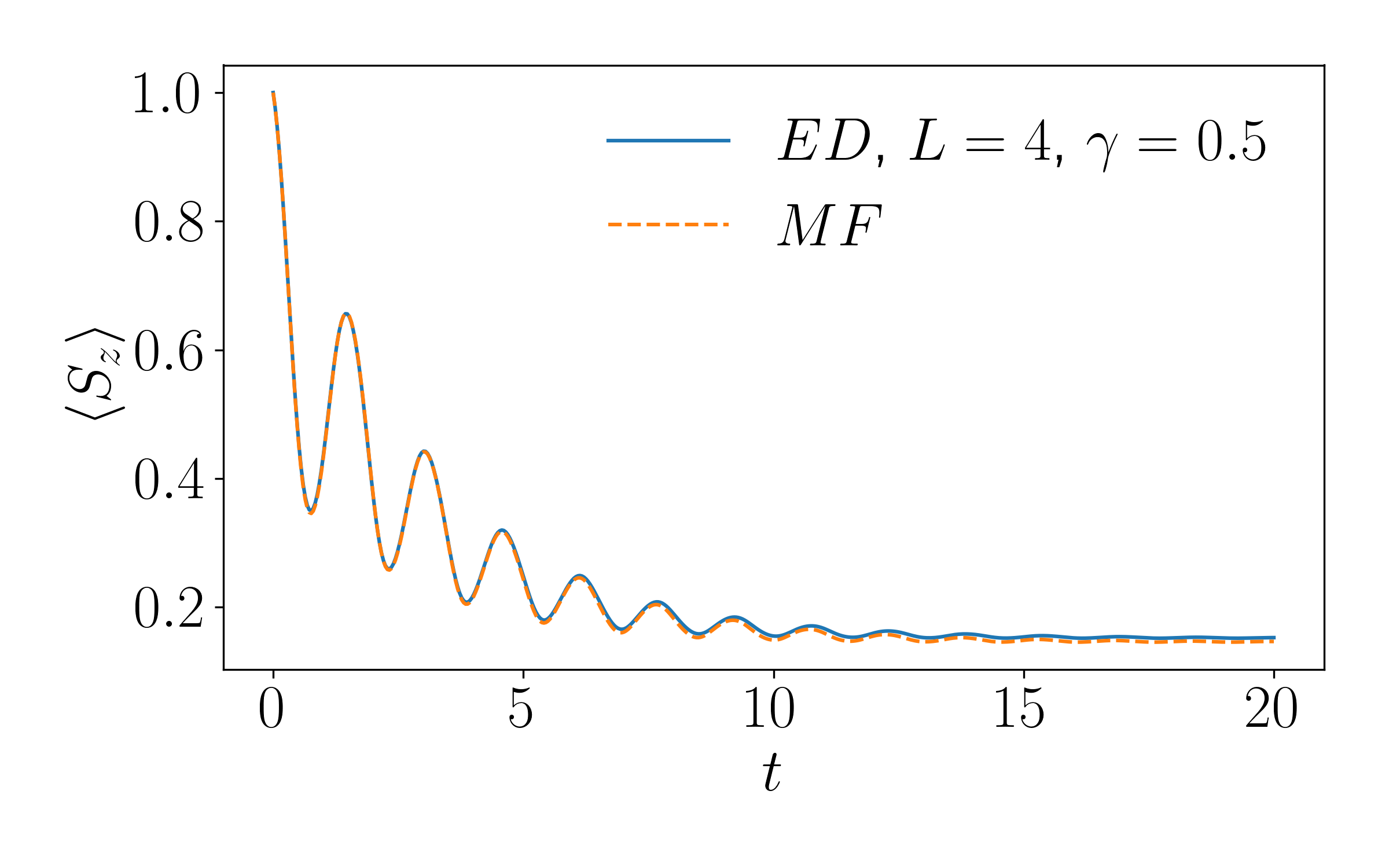}\put(-100,90){}
    \caption{\textbf{Spin-one model with fully-connected interaction}. Comparison of the dynamics of $S_z$ obtained from the mean-field equations in Eq.\ref{molla:eqn} and from the Lindblad master equation for the density matrix. The system is initialized with $\langle S_z\rangle = 1$ ($\lambda_7 = 1/\sqrt{2}$, $\lambda_8 = 1/\sqrt{6}$, all other $\lambda$’s set to zero). Numerical parameters: $\Omega = 3.0$, $\Delta = -7.0$, $E = 4.0$, $\chi = 1.0$.}
    \label{fig:Long-range_spinone}
\end{figure}
\begin{align}
A_{1j}
&=
\frac{\Omega}{2}\lambda_{5j}
+
(\Delta_j+E)\lambda_{4j}
-
\frac{\gamma}{2}\lambda_{1j},
\nonumber\\
A_{2j}
&=
\frac{\Omega}{2}
\left(
\lambda_{4j}
-
\lambda_{6j}
\right)
+
2E\lambda_{5j}
-
\gamma\lambda_{2j},
\nonumber\\
A_{3j}
&=
-\frac{\Omega}{2}\lambda_{5j}
-
(\Delta_j-E)\lambda_{6j}
-
\frac{\gamma}{2}\lambda_{3j},
\nonumber\\
A_{4j}
&=
-\Omega
\left(
\lambda_{7j}
+
\frac12\lambda_{2j}
\right)
-
(\Delta_j+E)\lambda_{1j}
-
\frac{\gamma}{2}\lambda_{4j},
\nonumber\\
A_{5j}
&=
-\frac{\Omega}{2}
\left(
\lambda_{1j}
-
\lambda_{3j}
\right)
-
2E\lambda_{2j}
-
\gamma\lambda_{5j},
\nonumber\\
A_{6j}
&=
\Omega
\left(
\frac12\lambda_{2j}
-
\frac{\sqrt{3}}{2}\lambda_{8j}
+
\frac12\lambda_{7j}
\right)
+
(\Delta_j-E)\lambda_{3j}
-
\frac{\gamma}{2}\lambda_{6j},
\nonumber\\
A_{7j}
&=
\Omega\lambda_{4j}
-
\frac{\Omega}{2}\lambda_{6j}
-
\frac{\gamma}{\sqrt{2}}
-
\gamma\lambda_{7j},
\nonumber\\
A_{8j}
&=
\frac{\sqrt{3}}{2}\Omega\lambda_{6j}
+
\frac{\gamma}{\sqrt{6}}
-
\gamma\lambda_{8j}.
\label{eq:spinone_A_terms}
\end{align}
The multiplicative-noise terms are
\begin{align}
D_{1j}
&=
\lambda_{7j}\xi_{+j}^{(1)}
+
\frac12
\left(
\lambda_{2j}\xi_{-j}^{(1)}
+
\lambda_{5j}\xi_{-j}^{(2)}
\right),
\nonumber\\
D_{2j}
&=
\frac12
\left[
\lambda_{3j}\xi_{+j}^{(1)}
+
\lambda_{6j}\xi_{+j}^{(2)}
+
\lambda_{1j}\xi_{-j}^{(1)}
-
\lambda_{4j}\xi_{-j}^{(2)}
\right],
\nonumber\\
D_{3j}
&=
\frac12
\left[
\lambda_{2j}\xi_{+j}^{(1)}
-
\lambda_{5j}\xi_{+j}^{(2)}
-
q_j\xi_{-j}^{(1)}
\right],
\nonumber\\
D_{4j}
&=
-\lambda_{7j}\xi_{+j}^{(2)}
+
\frac12
\left(
\lambda_{5j}\xi_{-j}^{(1)}
-
\lambda_{2j}\xi_{-j}^{(2)}
\right),
\nonumber\\
D_{5j}
&=
-\frac12
\left[
\lambda_{6j}\xi_{+j}^{(1)}
-
\lambda_{3j}\xi_{+j}^{(2)}
+
\lambda_{4j}\xi_{-j}^{(1)}
+
\lambda_{1j}\xi_{-j}^{(2)}
\right],
\nonumber\\
D_{6j}
&=
\frac12
\left[
\lambda_{5j}\xi_{+j}^{(1)}
+
\lambda_{2j}\xi_{+j}^{(2)}
-
q_j\xi_{-j}^{(2)}
\right],
\nonumber\\
D_{7j}
&=
-
\left[
\lambda_{1j}\xi_{+j}^{(1)}
-
\lambda_{4j}\xi_{+j}^{(2)}
+
\frac12
\left(
\lambda_{3j}\xi_{-j}^{(1)}
+
\lambda_{6j}\xi_{-j}^{(2)}
\right)
\right],
\nonumber\\
D_{8j}
&=
\sqrt{\frac{3}{2}}
\left[
\lambda_{3j}\xi_{-j}^{(1)}
+
\lambda_{6j}\xi_{-j}^{(2)}
\right].
\label{eq:spinone_D_terms}
\end{align}
Equations~\eqref{eq:spinone_A_terms} and~\eqref{eq:spinone_D_terms} give the
explicit drift and noise terms appearing in
Eq.~\eqref{eq:spinone_sde_compact}.
%
%
\section{Dynamics in the spin 1 model with $\alpha=0$}
\label{infinite:app}%
We can consider Eq.~\eqref{eq:spinone_hamiltonian} taking the $\alpha=0$ infinite-range Hamiltonian
\begin{equation}\label{oca1:eqn}
    \hat{H} = \frac{\Omega}{\sqrt{2}}\sum_{i=1}^L s_i^x - \Delta \sum_{i=1}^L\hat{n}_i-E\sum_i\hat{s}_i^z-\frac{\chi}{2L}\sum_{i\,,j} \hat{n}_i\hat{n}_j\,.
\end{equation}
We consider the dynamics of collective operators of the form
\begin{equation}
    \hat{\Lambda}_{\mu\,c} = \sum_j\hat{\Lambda}_{\mu\,j}\,.
\end{equation}
In order to study the dynamics of these operators we use the Lindblad Heisenberg equations~\cite{iemini-prl-2018:boundary-time-crystals}
\begin{equation}\label{Heis:eqn}
   \dot{\hat{\Lambda}}_{\mu\,c} = -i[\hat{H},\hat{\Lambda}_{\mu\,c}] +\frac{1}{2} \sum_{j\,\sigma}\left\{[\hat{L}_{j\,\sigma}^{{\dagger}},\hat{\Lambda}_{\mu\,c}]\hat{L}_{j\,\sigma}+\hat{L}_{j\,\sigma}^{{\dagger}}[\hat{\Lambda}_{\mu\,c},\hat{L}_{j\,\sigma}]\right\}\,.
\end{equation}
Using the commutation/anticommutation relations [see table~\ref{tab:gellmann-algebra}] we can therefore evaluate the Heisenberg equations Eq.~\eqref{Heis:eqn} as
\begin{align}\label{molla:eqn}
    \dot{\hat{\Lambda}}_{1\,c} &=  \frac{\Omega}{2}\hat{\Lambda}_{5\,c}+ (\Delta+E) \hat{\Lambda}_{4\,c} + \frac{\chi}{2L}\sum_{i,j}\{\hat{n}_i,\hat{\Lambda}_{4\,j}\}-\frac{\gamma}{2}\hat{\Lambda}_{1\,c}\nonumber\\
    \dot{\hat{\Lambda}}_{2\,c} &= \frac{\Omega}{2}\left(\hat{\Lambda}_{4\,c}-\hat{\Lambda}_{6\,c}\right)+2E\hat{\Lambda}_{5\,c}-{\gamma}\hat{\Lambda}_{2\,c}\nonumber\\
    \dot{\hat{\Lambda}}_{3\,c}& = -\frac{\Omega}{2}\hat{\Lambda}_{5\,c}-\left(\Delta - E\right)\hat{\Lambda}_{6\,c}-\frac{\chi}{2L}\sum_{i,j}\{\hat{n}_i,\hat{\Lambda}_{6\,j}\} -\frac{\gamma}{2}\hat{\Lambda}_{3\,c}\nonumber\\
    \dot{\hat{\Lambda}}_{4\,c}& = -\Omega\left(\hat{\Lambda}_{7\,c}+\frac12\hat{\Lambda}_{2\,c}\right) -(\Delta + E)\hat{\Lambda}_{1\,c} \nonumber\\
    &-\frac{\chi}{2L}\sum_{i,j}\{\hat{n}_i,\hat{\Lambda}_{1\,j}\}-\frac{\gamma}{2}\hat{\Lambda}_{4\,c}\nonumber\\
    \dot{\hat{\Lambda}}_{5\,c}& = -\frac{\Omega}{2}\left(\hat{\Lambda}_{1\,c}-\hat{\Lambda}_{3\,c}\right)-2E\hat{\Lambda}_{2\,c}-\gamma\hat{\Lambda}_{5\,c}\nonumber\\
    \dot{\hat{\Lambda}}_{6\,c}& = \Omega\left(\frac12\hat{\Lambda}_{2\,c}-\frac{\sqrt{3}}{2}\hat{\Lambda}_{8\,c}+\frac12\hat{\Lambda}_{7\,c}\right)+(\Delta -E)\hat{\Lambda}_{3\,c}\nonumber\\
    &+ \frac{\chi}{2L}\sum_{i,j}\{\hat{n}_i,\hat{\Lambda}_{3\,j}\}-\frac{\gamma}{2}\hat{\Lambda}_{6\,c}\nonumber\\
    \dot{\hat{\Lambda}}_{7\,c}& = \Omega\hat{\Lambda}_{4\,c}-\frac{\Omega}{2}\hat{\Lambda}_{6\,c}-\gamma\hat{\Lambda}_{7\,c}-\frac{\gamma}{\sqrt{2}}\hat{\boldsymbol{1}}_{c}\nonumber\\
    \dot{\hat{\Lambda}}_{8\,c}& = \frac{\sqrt{3}}{2}\Omega\hat{\Lambda}_{6\,c}-\gamma\hat{\Lambda}_{8\,c}+\frac{\gamma}{\sqrt{6}}\hat{\boldsymbol{1}}_{c}\,.
\end{align}
Considering the variables $\hat{\lambda}_\mu = \hat{\Lambda}_{\mu\,c} / L$, one sees that their commutator given by table~\ref{tab:gellmann-algebra} is depressed by a factor $1/L$ (an in this sense the effective Planck constant is order $1/L$). In particular in the thermodynamic limit $L\to\infty$ they commute, and one can neglect the quantum correlations between them, so that in this limit $\braket{\hat{\lambda}_\mu\hat{\lambda}_\nu} = \braket{\hat{\lambda}_\mu}\braket{\hat{\lambda}_\nu}$ up to corrections of order $1/L$ that vanish. (Here $\braket{\ldots}=\Tr[(\ldots) \hat{\rho}]$. The argument is very similar to the one used for a collective spin 1/2 model in~\cite{iemini-prl-2018:boundary-time-crystals}. Therefore, defining $\lambda_\mu(t) = \lim_{L\to\infty}\Tr[\hat{\Lambda}_{\mu\,c}\hat{\rho}(t)] / L$ one gets
%
\begin{align}\label{molla:eqn}
    \dot{\lambda}_{1} &=  \frac{\Omega}{2}\lambda_{5}+ (\Delta+E) \lambda_{4} \nonumber\\
    &+ \chi\lambda_4\left(\frac23 -\frac{\sqrt{6}}{6}\lambda_8+\frac{\sqrt{2}}{2}\lambda_7\right)-\frac{\gamma}{2}\lambda_1\nonumber\\
    \dot{\lambda}_{2} &= \frac{\Omega}{2}\left(\lambda_{4}-\lambda_{6}\right)+2E\lambda_{5}-{\gamma}\lambda_2\nonumber\\
    \dot{\lambda}_{3}& = -\frac{\Omega}{2}\lambda_{5}-\left(\Delta - E\right)\lambda_{6}\nonumber\\&-\chi\lambda_6\left(\frac23 -\frac{\sqrt{6}}{6}\lambda_8+\frac{\sqrt{2}}{2}\lambda_7\right) - \frac{\gamma}{2}\lambda_3\nonumber\\
    \dot{\lambda}_{4}& = -\Omega\left(\lambda_{7}+\frac12\lambda_{2}\right) -(\Delta + E)\lambda_{1} \nonumber\\&-\chi\lambda_1\left(\frac23 -\frac{\sqrt{6}}{6}\lambda_8+\frac{\sqrt{2}}{2}\lambda_7\right)-\frac{\gamma}{2}\lambda_{4}\nonumber\\
    \dot{\lambda}_{5}& = -\frac{\Omega}{2}\left(\lambda_{1}-\lambda_{3}\right)-2E\lambda_{2}-\gamma\lambda_{5}\nonumber\\
    \dot{\lambda}_{6}& = \Omega\left(\frac12\lambda_{2}-\frac{\sqrt{3}}{2}\lambda_{8}+\frac12\lambda_{7}\right)+(\Delta -E)\lambda_{3}\nonumber\\&+ \chi\lambda_3\left(\frac23 -\frac{\sqrt{6}}{6}\lambda_8+\frac{\sqrt{2}}{2}\lambda_7\right)-\frac{\gamma}{2}\lambda_{6}\nonumber\\
    \dot{\lambda}_{7}& = \Omega\lambda_{4}-\frac{\Omega}{2}\lambda_{6}-\gamma{\lambda}_{7\,c}-\frac{\gamma}{\sqrt{2}}\nonumber\\
    \dot{\lambda}_{8}& = \frac{\sqrt{3}}{2}\Omega\lambda_{6}-\gamma{\lambda}_{8\,c}+\frac{\gamma}{\sqrt{6}}\,.\nonumber\\
\end{align}

%

Notice that these equations coincide with the {semiclassical Langevin} ones Eq.~\eqref{eq:spinone_sde_appendix} when noise terms $D_{\mu j}$ are neglected. In Fig.~\ref{fig:Long-range_spinone}, we show the dynamics of $S_z$ obtained by solving the above equation. The system is initialized in a state with $\langle S_z\rangle = 1$, corresponding to $\lambda_7 = 1/\sqrt{2}$, $\lambda_8 = 1/\sqrt{6}$, and all other $\lambda$’s set to zero. The results agree with those obtained from solving the Lindblad master equation for the density matrix -- the agreement being valid up to the Ehrenfest time $t^*(L)$ discussed in Sec.~\ref{devi:sec} -- and we have also checked that exactly coincide with the results of the mean-field scheme developed in~\cite{Wang2025BoundaryTimeCrystals}.

\end{document}